\documentclass[journal]{IEEEtran}

\ifCLASSINFOpdf
  \usepackage[pdftex]{graphicx}
\else
\fi
\ifCLASSOPTIONcompsoc
 \usepackage[caption=false,font=normalsize,labelfont=sf,textfont=sf,margin=5pt]{subfig}
\else
 \usepackage[caption=false,font=footnotesize,margin=10pt]{subfig}
\fi

\usepackage[english]{babel}
\usepackage{amsmath}
\usepackage{amssymb}
\usepackage{centernot}
\usepackage{algorithm}
\usepackage{algorithmic}
\usepackage{tikz}
\usepackage{pgfplots}
\usepackage{url}
\usepackage{color}
\usepackage{soul}
\usepackage[normalem]{ulem}
\usepackage{stmaryrd}
\usepackage{subfig}
\usepackage{ulem}
\usepackage{pdfpages}

\usetikzlibrary{calc,shapes,positioning,plotmarks}

\floatname{algorithm}{{A}lgorithm}

\newcommand{\floor}[1]{\left \lfloor #1 \right \rfloor}
\newcommand{\ceil}[1]{\left \lceil #1 \right \rceil}
\renewcommand{\vec}[1]{\mathbf{#1}}
\newcommand{\mat}[1]{\mathbf{#1}}
\newcommand{\norm}[2]{\lVert #1 \rVert_{#2}}

\DeclareMathOperator*{\argmin}{arg\,min}

\newcommand{\dotprod}[2]{\left\langle#1, #2\right\rangle}

\renewcommand\sout{\bgroup\markoverwith
{\textcolor{red}{\rule[.5ex]{2pt}{0.8pt}}}\ULon}

\newtheorem{theorem}{Theorem}

\begin{document}

\title{Adaptive Hierarchical Sensing for the Efficient Sampling of Sparse and Compressible Signals}

\author{Henry~Sch\"utze,
        Erhardt~Barth,~\IEEEmembership{Member,~IEEE,}
        and~Thomas~Martinetz,~\IEEEmembership{Senior~Member,~IEEE}
\thanks{The authors are with the Institute for Neuro- and Bioinformatics, University of L\"ubeck, Ratzeburger Allee 160, 23562 L\"ubeck, Germany}
}


\markboth{}%
{}

\maketitle

\begin{abstract}
 We present the novel adaptive hierarchical sensing algorithm $K$-AHS, which samples sparse or compressible signals with a measurement complexity equal to that of Compressed Sensing (CS). In contrast to CS, $K$-AHS is adaptive as sensing vectors are selected while sampling, depending on previous measurements. Prior to sampling, the user chooses a transform domain in which the signal of interest is sparse. The corresponding transform determines the collection of sensing vectors. $K$-AHS gradually refines initial coarse measurements to significant signal coefficients in the sparse transform domain based on a sensing tree which provides a natural hierarchy of sensing vectors. $K$-AHS directly provides significant signal coefficients in the sparse transform domain and does not require a reconstruction stage based on inverse optimization. Therefore, the $K$-AHS sensing vectors must not satisfy any incoherence or restricted isometry property. A mathematical analysis proves the sampling complexity of $K$-AHS as well as a general and sufficient condition for sampling the optimal $k$-term approximation, which is applied to particular signal models. The analytical findings are supported by simulations with synthetic signals and real world images. On standard benchmark images, $K$-AHS achieves lower reconstruction errors than CS.
\end{abstract}


\IEEEpeerreviewmaketitle

\section{Introduction}

\IEEEPARstart{D}{uring} the last decade Compressed Sensing (CS) has rapidly emerged and is now established as a useful sampling technique in various engineering disciplines. Many digital acquisition devices, for instance digital cameras, first fully sample the analog signal of interest and subsequently perform lossy compression to get rid of the vast amount of redundant information collected in the first stage. CS, on the contrary, is a much more efficient approach as it embeds the data compression step into the sampling stage. Given the signal is sparse or compressible in some transform domain, the total number of CS measurements is much lower than the Nyquist-Shannon sampling theorem demands for classical sampling. The sparseness assumption holds for many natural signal classes. Classical sampling of a signal of interest, e.g. a visual scene, can be seen as making linear measurements in terms of inner products of the signal with canonical basis functions. With CS, inner products of the signal are instead measured with alternative (e.g. random) functions. Given the set of collected measurements, the signal is reconstructed by solving a convex $\ell_1$-norm optimization problem or by using a greedy $\ell_0$-norm pursuit method. CS has found widespread applications, ranging from radar imaging \cite{PoErPaCe10, Ender2010} over Magnetic Resonance Imaging (MRI) \cite{LuDoSaPa07} to one pixel cameras \cite{TaLaWaDuBaSaKeBa06, WaLaDuBaSaTaKeBa06, WaLaDuBaSaTaKeBa06b, WeEdBoJoSuPa13, SuEdBoViWeBoPa13}.

In this paper we present an alternative approach, where sensing vectors are selected dependent on values of previously observed measurements. In this sense our approach is adaptive. Adaptive sensing schemes have been proposed before. For example, Coulter at al. proposed the neural network model Adaptive Compressed Sensing (ACS), which is a sparse coding neural network with a synaptic learning scheme that is embedded into the compressed sensing framework. Motivated by neuro\-biological findings, encoding and weight adaptation stages of their ACS network have limited access to the original data. They showed that with these networks smooth and biologically realistic receptive fields, also known from sparse coding models, emerge despite the fact that the sensory input is subsampled and mixed by the feedforward connectivity \cite{CoHiIsSo10}.

Burciu et al. proposed Hierarchical Manifold Sensing (HMS), an adaptive hierarchical sensing scheme to solve classification tasks for images that are distributed on a non-linear manifold. By hierarchically decomposing the training data into partitions using PCA and k-means clustering, HMS infers the class of an input image based on only few linear measurements \cite{BuMaBa16a}. Their approach, however, has limitations as it requires to have instances in the training set which are similar to the unknown signal that is to be classified.

Adaptive sensing based on Bayesian inference has its roots in the area of experimental design \cite{HaNo12}, which addresses the problem to optimally design a sequence of experiments in order to gain knowledge about the true state of the world. The outcome of each experiment can reduce the experimenters uncertainty about the state by providing new bits of information. The experimenters objective is to exploit the information of previous experiments and design the subsequent experiment in a way that maximizes the expected information gain \cite{degroot62, lindley56}. Bayesian Adaptive Sensing is a framework which follows this concept of optimal design in order to sample an unknown sparse signal sequentially using multiple random sensing matrices. Entries of these sensing matrices are drawn from a symmetric distribution which is gradually adjusted over time taking observed measurements into account. This is in contrast to non-adaptive CS, where only a single sensing matrix is used whose entries are drawn i.i.d. from a symmetric distribution. The variance of the distribution is adaptively adjusted. Thus, sensing energy is focused onto locations for which it is rather likely that signal components are contained. For a new sensing step the sensing matrix is drawn from the distribution that maximizes the Kullback-Leibler divergence between the posterior distribution of the signal given the measurements and the prior distribution of the signal \cite{HaNo12}. Bayesian Adaptive Sensing can outperform non-adaptive CS in noisy settings in terms of the reconstruction error relative to the number of measurements \cite{CaHaNoRa08, JiXuCa08, Seeger08, SeNi08}.


Deutsch et al. proposed Adaptive Direct Sampling (ADS) to directly sample relevant wavelet coefficients of an image in a selective hierarchical manner \cite{DeAvDe09}. The set of possible sensing vectors matches with the wavelet basis. First, ADS samples all transform coefficients in all sub-bands within a limited number of the coarsest scales. Subsequently, a heuristic based on the Lipschitz exponent is applied to iteratively decide at which image locations and for which sub-bands the coefficients of the next finer scale will be sampled or omitted. Their approach, however, is limited to the wavelet domain.

Aldroubi et al. proposed an adaptive compressed sampling approach to sample sparse signals based on a Huffman tree \cite{AlWaZa11}. The Huffman tree is derived from probabilities assigned to sets of non-zero locations, which reflect statistics of the signal population. In a way, such a Huffman tree is related to the sensing tree that is used by K-AHS (see Section \ref{sec:sensing_tree} below) as it is traversed during sampling and each visited node corresponds to a linear measurement of the signal with a sensing vector that yields the sum of a subset of signal components. On average, their method has a sampling complexity of $k\log N + 2k$ measurements to find $k$ non-zero locations. In contrast to $K$-AHS, their sampling scheme traverses the Huffman tree multiple times (one run for each non-zero component), and requires furthermore to recalculate sensing vectors after each run, depending on already identified non-zero locations. However, the authors do not address the issue that, for compressible signals, unfavorable constellations of significant coefficients can cancel each other. Furthermore, their method was not tested on real world signals, or non-canonical sparse transform domains.

In \cite{ScBaMa14}, a threshold-based variant of AHS has been proposed to sample $k$-sparse signals by less than $2k (\log N/k+1)$ measurements. It is conceptually based on the same kind of sensing tree as will be introduced in Section \ref{sec:sensing_tree}. In contrast to $K$-AHS, the threshold parameter gives little control on the total number of measurements if the signal is compressible rather than strictly $k$-sparse. This limitation and the lack of a theoretical analysis motivate the novel sampling method $K$-AHS that we propose in this paper.

\subsection{Contribution and Structure of this Article}

In Section \ref{sec:K-AHS}, we introduce our novel adaptive hierarchical sensing algorithm $K$-AHS and the sensing tree it is based on, explain its reconstruction stage, and prove that its sampling complexity is of the order $\mathcal{O}(K \log N/K)$. Section \ref{sec:SSCI} is dedicated to the theoretical analysis of $K$-AHS in terms of recovering the most significant signal coefficients particularly for signals that are not strictly $k$-sparse but compressible. Our Theorem \ref{th2} states a general sufficient condition to guarantee the detection of the $k$ most significant signal coefficients. We use it to derive conditions on model parameters for three different signal models such that sensing by $K$-AHS will be successful. We illustrate sensing performance of $K$-AHS for synthetic signals and validate our theoretical findings. In Section \ref{sec:results} we use $K$-AHS for compressive imaging of real-world images and report reconstruction accuracy dependent on the number of measurements.

\section{The $K$-AHS Algorithm}
\label{sec:K-AHS}

\subsection{Prerequisite}

Assume that $\vec{x} \in \mathbb{R}^N$ is the unknown signal of interest. The main prerequisite for sensing with $K$-AHS is, that a linear basis $\mat{\Psi} \in \mathbb{R}^{N \times N}$ (analysis basis) is known that transforms (analysis transform) the signal $\vec{x}$ to a sparse representation $\vec{a} = \mat{\Psi}\vec{x}$ which has only few entries substantially different from zero. Let $\overline{\mat{\Psi}}^T \in \mathbb{R}^{N \times N}$ be the corresponding inverse linear basis (synthesis basis) that transforms (synthesis transform) the sparse representation $\vec{a}$ back to the original signal $\vec{x} = \overline{\mat{\Psi}}^T\vec{a}$. For instance, $\mat{\Psi}$ can be an orthogonal basis (with the implication $\overline{\mat{\Psi}} = \mat{\Psi}$) such as the Discrete Cosine Transform (DCT) or a Daubechies wavelet basis. Alternatively, the pair $\mat{\Psi}$ and $\overline{\mat{\Psi}}$ can be a biorthogonal basis such as a Cohen-Daubechies-Feauveau wavelet basis.

\subsection{Sensing Tree}
\label{sec:sensing_tree}
For now we assume that $N$ is a power of $2$. The key data structure underlying $K$-AHS is a so called sensing tree. It is a perfect binary tree of height $\log_2N$ with $2N-1$ nodes. Each node $(l,n)$ of the tree is associated with a sensing vector $\vec{\varphi}_{l,n}$, where $l=0, \dots, \log_2 N$ is the index of the tree level (starting at the bottom level), and $n=1,\dots,N2^{-l}$ is the index of the node within level $l$. 

The sensing vectors of the bottom level correspond to elements of analysis basis $\mat{\Psi} = \left[\vec{\psi}_1, \dots, \vec{\psi}_N\right]$ in which $\vec{x}$ is assumed to have a sparse representation, i.e.
\begin{eqnarray}
  \label{eq:sensing_vec_leaves}
  \vec{\varphi}_{0,n}=\vec{\psi}_n,\,n=1, \dots, N\,.
\end{eqnarray}
In a bottom-up manner, the sensing vector of each internal node is the sum of sensing vectors assigned to its two direct descendant nodes, i.e. for any $l\in\{1,...,\log_2N\}$
\begin{equation}
  \label{eq:sensing_vec_sum_direct_descendants}
  \vec{\varphi}_{l,n} =  \vec{\varphi}_{l-1, 2n-1} + \vec{\varphi}_{l-1, 2n} \quad, n=1, \dots, N2^{-l}\,.
\end{equation}

By construction, $\vec{\varphi}_{l,n}$ can also be written as the sum of a subset of basis vectors from $\mat{\Psi}$:
\begin{eqnarray}
  \label{eq:sensing_vec_sum_leaves}
  \vec{\varphi}_{l,n} &=& \sum_{i = (n-1)2^{l}+1}^{n2^{l}} \vec{\psi}_i\,.
\end{eqnarray}
The set of analysis basis vectors that forms $\vec{\varphi}_{l,n}$ corresponds to the leaves of the subtree with root node $(l,n)$. 

\begin{figure}[htb]
  \centering
  
  \begin{tikzpicture}[scale=0.5,
    node distance=1.5cm and 0.75cm,
    mynode/.style={
      fill=black,
      circle,
      inner sep=0pt,
      circle,
      minimum size=.2cm}
  ]

  \node[mynode] at ([yshift=-0.36\textheight]current page.north) (root) {};
  \node[mynode,below left=of root] (L) {};
  \node[mynode,below right=of root] (R) {};
  \node[mynode,below left=of L] (LL) {};
  \node[mynode,below right=of R] (RR) {};
  \node[mynode,below left=of LL] (LLL) {};
  \node[mynode,below right=of LL] (LLR) {};
  \node[mynode,below left=of RR] (RRL) {};
  \node[mynode,below right=of RR] (RRR) {};

  \coordinate[below right=of L] (LR);
  \coordinate[below left=of R] (RL);

  \path (L) -- coordinate[pos=0.3333] (aux1) coordinate[pos=0.6666] (aux2) (LL);
  \path (R) -- coordinate[pos=0.3333] (aux3) coordinate[pos=0.6666] (aux4) (RR);
  \path (L) -- coordinate[pos=0.3333] (aux5) (LR);
  \path (R) -- coordinate[pos=0.3333] (aux6) (RL);


  \draw (L) -- (root) -- (R);
  \draw (LLL) -- (LL) -- (LLR);
  \draw (RRL) -- (RR) -- (RRR);

  \draw[dash pattern=on 1pt off 2pt, thick] (L) -- (aux1);
  \draw[dash pattern=on 1pt off 2pt, thick] (aux2) -- (LL);
  \draw[dash pattern=on 1pt off 2pt, thick] (L) -- (aux5);
  \draw[dash pattern=on 1pt off 2pt, thick] (R) -- (aux3);
  \draw[dash pattern=on 1pt off 2pt, thick] (aux4) -- (RR);
  \draw[dash pattern=on 1pt off 2pt, thick] (R) -- (aux6);
  \draw[dash pattern=on 1pt off 2pt, thick] (L|-LL) -- (R|-RR);
  \draw[dash pattern=on 1pt off 2pt, thick] ([xshift=5pt]aux5|-LLR) -- ([xshift=-5pt]aux6|-RRL);

  \node[above=3pt of root.north]
    {$\vec{\varphi}_{\log_2N,1}$};

  \node[left=3pt of L,anchor=east]
    {$\vec{\varphi}_{\log_2\frac{N}{2},1}$};
  \node[right=3pt of R,anchor=west]
    {$\vec{\varphi}_{\log_2\frac{N}{2},2}$};

  \node[left=3pt of LL,anchor=east]
    {$\vec{\varphi}_{1,1}$};
  \node[right=3pt of RR,anchor=west]
    {$\vec{\varphi}_{1,\frac{N}{2}}$};

  \node[below=3pt of LLL]
    {$\underbrace{\vec{\varphi}_{0,1}}_{\vec{\psi}_1}$};
  \node[below=3pt of LLR]
    {$\underbrace{\vec{\varphi}_{0,2}}_{\vec{\psi}_2}$};
  \node[below=3pt of RRL]
    {$\underbrace{\vec{\varphi}_{0,N-1}}_{\vec{\psi}_{N-1}}$};
  \node[below=3pt of RRR]
    {$\underbrace{\vec{\varphi}_{0,N}}_{\vec{\psi}_N}$};
\end{tikzpicture}

  \caption{Schematic illustration of the $K$-AHS sensing tree. To each node $(l,n)$ a sensing vector $\vec{\varphi}_{l,n}$ is assigned. The first index $l \in \{0,...,\log_2N\}$ indicates the tree level starting with $l=0$ at the bottom level. The second index $n \in \{1,...,N2^{-l}\}$ is the node index for level $l$. There is a one to one matching between sensing vectors of leaf nodes and elements of analysis basis $\mat{\Psi}$.}
  \label{fig:sensing_tree_sketch}
\end{figure}
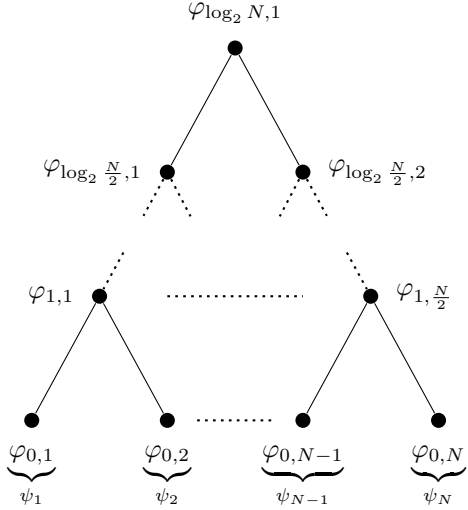

\figurename\ \ref{fig:sensing_tree_sketch} illustrates the sensing tree schematically.

\subsection{Sensing Procedure}

$K$-AHS selectively traverses the sensing tree, level by level. For each node $(l,n)$ that is visited, one linear measurement is collected by the sensing operation $\dotprod{\vec{x}}{\vec{\varphi}_{l,n}}$, i.e. by the inner product between the unknown signal of interest $\vec{x}$ and the node specific sensing vector $\vec{\varphi}_{l,n}$. Note that due to property (\ref{eq:sensing_vec_sum_leaves}), and the bilinearity of the inner product for real vector spaces, a sensing operation implicitly calculates the sum of signal coefficients in the sparse transform domain $\mat{\Psi}$, i.e.
\begin{eqnarray}
  \label{eq:sum_of_coefficients}
  \dotprod{\vec{x}}{\vec{\varphi}_{l,n}} &=& \sum_{i = (n-1)2^{l}+1}^{n2^{l}} a_i\,.
\end{eqnarray}

The $K$-AHS algorithm has a user parameter $K$. This parameter allows to control how many signal coefficients are going to be identified. Furthermore, it determines how many nodes $K$-AHS takes into consideration when it transitions from one level to the next. The direct descendants of the nodes corresponding to the $K$ largest\footnote{In the following, for $K$-AHS measurements the relation \textit{larger} and \textit{smaller} is exclusively meant in terms of their magnitude.} measurements are visited in the next iteration. Thus, there are $2K$ sensing operations for the new level from which again the nodes coinciding with the $K$ largest measurements are further processed. This iterative scheme is continued until $2K$ leaf nodes are reached. The $K$-AHS pseudo code is listed in Algorithm \ref{AHS_pseudocode}. The idea is that, by this procedure, the $K$ largest entries of $\vec{a} = \mat{\Psi}\vec{x}$ are collected. For instance, if $\vec{a}$ has at most $K$ non-zero entries (without any subset summing up to exactly zero, e.g., when drawn from a continuous probability distribution), then the signal is completely sensed and can be perfectly reconstructed. In Section \ref{sec:SSCI} we investigate further models of compressive signals.

\begin{algorithm}
  \footnotesize
  \caption{Adaptive Hierarchical Sensing ($K$-AHS)}
  \label{AHS_pseudocode}
  \begin{algorithmic}[1]
  \REQUIRE{``Interface'' to collect inner products $\dotprod{\vec{x}}{\cdot}$ with signal $\vec{x} \in \mathbb{R}^N$\\
  Invertible transform $\mat{\Psi} \in \mathbb{R}^{N\times N}$ yielding sparse representation $\vec{a} = \mat{\Psi}\vec{x}$\\
  Target sparsity level $K < \frac{N}{4}$}
  \ENSURE{Sparse estimate $\vec{\hat{a}}$ of $\vec{a}$, where $\norm{\vec{\hat{a}}}{0} \leq 2K$}
  \STATE{Set initial sensing tree level $L$ according to (\ref{eq:initial_level})}
  \STATE{Collect all measurements of level $L$
  \begin{equation}
    \nonumber
    \dotprod{\vec{x}}{\vec{\varphi}_{L,n}},\,n=1,\dots,N2^{-L}
  \end{equation}}
  \FOR{$l = L,L-1,..., 1$}
    \STATE{Let $n_1,...,n_K$ be the subscripts of the $K$ largest measurements collected from sensing tree level $l$}
    \STATE{At the next level $l-1$, collect for $j=1,...,K$ the measurements of the two child nodes of node $(l,n_j)$
    \begin{equation}
      \nonumber
      \dotprod{\vec{x}}{\vec{\varphi}_{l-1,2n_j-1}}\,, \quad \dotprod{\vec{x}}{\vec{\varphi}_{l-1,2n_j}}
    \end{equation}
    }
  \ENDFOR
  \STATE{For $n=1, \dots, N$, set
    \begin{equation}
      \nonumber
      \hat{a}_n \leftarrow 
      \begin{cases}
	 \dotprod{\vec{x}}{\vec{\varphi}_{0,n}} & \text{if observed}\\
	0 & \text{otherwise}
      \end{cases}
    \end{equation}}
  \end{algorithmic}
\end{algorithm}

In order to avoid unnecessary sensing operations, $K$-AHS does not start with the first measurement at the root node of the sensing tree but at a suitable initial level $L$. This initial level has to be sensed completely in order to identify the $K$ nodes providing the largest measurements for processing the next level. At each subsequent level $l < L$, only $2K$ measurements are collected. Regarding the total number of measurements the optimal initial tree level depends on the user parameter $K$ and is given by 
\begin{equation}
  \label{eq:initial_level}
  L = \log_2N - \left\lfloor \log_2K \right\rfloor - 2\,.
\end{equation}
$L$ is the highest level $l \in \{0,\dots,\log_2N/4\}$ that contains more than $2K$ nodes. For example, for $K=1$ we start with the level $L=\log_2N/4$ which contains $4$ nodes. For $N/4 \leq K \leq N/2$ we obtain $L=0$ and $N$ measurements, a trivial scenario where each coefficient is sensed individually. This shows that K-AHS makes sense only for small values of $K$, i.e., sparse signals. 

The sensing scheme provided by $K$-AHS is adaptive as each measurement collected in level $l < L$ depends on large measurements at the corresponding ancestor nodes. Furthermore, $K$-AHS operates hierarchically as the transition from a node in level $l$ to its two child nodes in level $l-1$ splits the partition of coefficients, which are summed up, in two halves. In that sense, the sensing scheme can be seen as a successive refinement of initially coarse measurements up to a set of significant signal coefficients in the sparse transform domain.

\subsection{Signal Reconstruction}

Due to Eq. (\ref{eq:sensing_vec_leaves}), $K$-AHS directly senses $2K$ entries of $\vec{a}$ at the bottom level of the sensing tree. These sensed coefficients are used to built $\vec{\hat{a}} \in \mathbb{R}^N$ as the estimation of $\vec{a}$. The remaining $N-2K$ entries of $\vec{\hat{a}}$ are set to zero. The reconstructed signal in the original domain $\vec{\hat{x}}$ is easily obtained by applying the synthesis transform
\begin{equation}
  \label{eq:signal_reconstruction}
  \vec{\hat{x}} = \overline{\mat{\Psi}}^T\vec{\hat{a}}\,.
\end{equation}
Note that $K$-AHS differs in an important point from CS as no inverse optimization problem has to be solved to obtain $\vec{\hat{a}}$.

Why do we expect to obtain most of the signal energy with this kind of hierarchical sensing? It is easy to see that exactly K-sparse signals are sensed perfectly by K-AHS (under the mild assumption that no subsets of non-zero coefficients sum up to zero), i.e., the $2K$ coefficients of the final step contain all the non-zero elements and $\vec{\hat{a}}=\vec{a}$. In case the signal is not exactly K-sparse, it might happen that significant coefficients cancel out themselves within a measurement sum and get lost. However, if the coefficients are drawn from a heavy-tailed distribution, which is a characteristic property of sparse natural signals, for two coefficients $a_i$, $a_j$ 
\begin{equation}
  \lim_{c\rightarrow\infty} \mathrm{Prob}\,\left[|a_i-a_j|<t\mid a_i,a_j>c\right] = 0 \; \hbox{for all}\; t>0
\end{equation}
is valid. This property of heavy-tailed distributed random variables implies that the probability, that two large coefficients sum up to a small value and get lost during the sensing process, converges to zero the more significant these coefficients are. This is the basic idea behind K-AHS. Indeed, in Section \ref{sec:results} we will see that natural images can successfully be sensed by K-AHS. In Section \ref{sec:application_th2_k-sparse_model}, we will analyze the sensing quality of K-AHS more rigorously. 

\subsection{Sampling Complexity}

$K$-AHS has a sampling complexity of the same order as Compressed Sensing.
\begin{theorem}
  \label{theorem_1}
  Let $\vec{x} \in \mathbb{R}^N$ and $1\leq K < N/4$. For $M$, the total number of $K$-AHS measurements, the following bound holds
  \begin{equation}
  \label{eq:bound_on_M}
   M \leq 2K\log_2\frac{N}{K}\,.
  \end{equation}
\end{theorem}
\begin{IEEEproof}
  According to Algorithm \ref{AHS_pseudocode}, $K$-AHS entirely processes the initial level $L$ of the sensing tree, which results in $N2^{-L}$ measurements. There are $L$ subsequent levels, each adds $2K$ measurements. Hence,
  \begin{eqnarray}
    \label{eq:exact_number_AHS_measurements}
    M &=& N2^{-L} + 2KL\,.
  \end{eqnarray}
  Plugging (\ref{eq:initial_level}) into (\ref{eq:exact_number_AHS_measurements}) yields
  \begin{eqnarray}
    \nonumber
    M	&=& 2^{\floor{\log_2K} + 2} + 2K(\log_2N - \floor{\log_2K} - 2)\\
	\label{eq:M_ineq_bound}
	&\leq& 2^{\log_2K  + 2} + 2K(\log_2 N - \log_2K -2)\\
	\nonumber
	&\leq& 2K \log_2\frac{N}{K}\,.
  \end{eqnarray}
  For (\ref{eq:M_ineq_bound}), we have used the inequality
  \begin{equation*}
    2^{\floor{\log_2K}+2} - 2K\floor{\log_2K} \leq 2^{\log_2K+2} - 2K\log_2K\,.
  \end{equation*}
\end{IEEEproof}
Equality in (\ref{eq:bound_on_M}) holds if $K \in \{1,2,4,8,\dots\}$.

\subsection{Extension to Arbitrary $N$ and Weighted Sums}
\label{sec:extension_arbitrary_N_and_weighted_sums}
To handle a signal dimensionality $N$ which is not a power of base 2, we simply expand the analysis basis $\mat{\Psi}$ by $\tilde{N}-N$ additional zero elements, where $\tilde{N} = 2^{\ceil{\log_2N}}$. The expanded analysis basis $\tilde{\mat{\Psi}} \in \mathbb{R}^{N \times \tilde{N}}$ is then given by
\begin{equation*}
 \tilde{\mat{\Psi}} = \left[ \mat{\Psi}, \mat{0}_{N \times (\tilde{N}-N)} \right]\,.
\end{equation*}
The size of the sensing tree will be increased due to the additional artificial sensing vectors. However, most of these additional nodes will be discarded very early during sensing as they provide merely zero measurements. For the reconstruction, only the $N$ original dimensions will be used. The artificial $\tilde{N}-N$ components of $\hat{\vec{a}}$ will be zero and can be cut off such that the original synthesis transform $\overline{\mat{\Psi}}$ is used for the reconstruction as stated in Eq. (\ref{eq:signal_reconstruction}).

According to (\ref{eq:sensing_vec_sum_direct_descendants}), the sensing vector of an internal node of the sensing tree is constructed by the sum of the sensing vectors assigned to its direct descendant nodes. It might be suitable to generalize (\ref{eq:sensing_vec_sum_direct_descendants}) such that the direct sum becomes a weighted sum:
\begin{eqnarray}
  \label{eq:sensing_vec_linear_combi_I}
  \vec{\varphi}_{l,n} &=& \alpha_{l,n}\, \vec{\varphi}_{l-1, 2n-1} + \beta_{l,n}\,\vec{\varphi}_{l-1, 2n}\,,
\end{eqnarray}
where $\alpha_{l,n}$ and $\beta_{l,n}$ are real non-zero weights.
This can be useful if the signal class of interest has particular statistical properties. For instance, when the measurements provided by two sibling nodes $\vec{\varphi}_{l, 2n-1}$ and $\vec{\varphi}_{l, 2n}$ are strongly anti-correlated, it would be advantageous to choose weights $\alpha_{l,n}$ and $\beta_{l,n}$ with opposite signs.

\section{Sensing Quality}
\label{sec:SSCI}

When a signal $\vec{x}$ is sampled by $K$-AHS, the optimal result one can expect is that the $K$ largest entries of its coefficient vector $\vec{a}$ are collected. Whether this optimal result can be achieved depends on the compressibility of the signal. In the following, we will introduce three signal models for which we analyze the sensing performance of $K$-AHS. 

\subsection{Signal Models}
\label{subsec:signal_models}

The following signal models characterize the decay of signal coefficients. Let $h_1,...,h_n$ be a sequence of indices which sorts the entries of $\vec{a}$ in descending order of their magnitudes, i.e., $|a_{h_1}| \geq |a_{h_2}| \geq ... \geq |a_{h_N}|$. 
Each signal model assumes certain properties regarding $|a_{h_n}|$, $n = 1,\dots,N$.

\subsubsection{$k$-Sparse Model}

\label{sec:l0-sparse_signal_model}

A $k$-sparse signal, denoted by $\norm{\vec{x}}{0}=k$, has the property
\begin{eqnarray}
  |a _{h_n}|
  \begin{cases}
    > 0,& \text{if } n \leq k\\
    = 0, & \text{otherwise}\,.
  \end{cases}
\end{eqnarray}
Commonly, the number of non-zero coefficients is very small compared to the signal dimensionality, i.e. $k \ll N$. We furthermore assume that the $k$ non-zero coefficients come from a continuous probability distribution.

\subsubsection{Exponential Model}
\label{sec:exp_decay_signal_model}

The decay of the coefficient magnitudes can be modeled by an exponential law
\begin{eqnarray}
  |a _{h_n}| = Rq^{-n+1}\,,
\end{eqnarray}
where base $q>1$ is the model parameter and $R>0$ is a scaling constant.

\subsubsection{Power Law Model}
\label{sec:power_law_decay_signal_model}

Similar to \cite{CandesTao06}, the decay of the coefficient magnitudes can be modeled by a power law
\begin{eqnarray}
  |a _{h_n}| = R n^{-\alpha}\,,
\end{eqnarray}
where exponent $\alpha > 1$ is the model parameter and $R>0$ is a scaling constant. It has been shown that many natural signal classes are consistent with this model \cite{CandesTao06,DeVore98,DoVeDeDa98,Mallat08}.

\subsection{Sufficient Condition to Collect the $k$ largest Coefficients}

Due to (\ref{eq:sum_of_coefficients}), a sensing operation $\dotprod{\vec{x}}{\vec{\varphi}_{l,n}}$ implicitly calculates the sum of a partition of $\vec{a}$. Similarly, any sensing operation $\dotprod{\vec{x}}{\vec{\varphi}_{l,n'}}$ at any other node $(l,n')$ of the same level calculates a sum of another disjoint partition of $\vec{a}$. For any tree level $l$, the size of such a partition (number of summands) is $2^l$. Merely the $K$ nodes with the largest measurements (the largest sums) are further processed. Consequently, the absolute value of those measurements, which include significant coefficients, should not become too small. In particular, significant coefficients should not cancel each other within a sum.

Let ${\cal K}=\{ a_{h_1}, \dots, a_{h_k}\}$ be the set of the $k$ largest coefficients we want to collect (we call them significant coefficients). We define $u$ as the smallest absolute value that can occur by summing up a subset of these significant coefficients, i.e.
\begin{equation}
  \label{eq:smallest_value_s}
  u=\min_{{\cal A}\subseteq {\cal K}} \left | \sum_{a_n\in {\cal A}} a_n \right | \, .
\end{equation}
The following theorem states a sufficient condition for $K$-AHS to be successful in collecting all significant coefficients.
\begin{theorem}
  \label{th2}
  Let $k\leq K$, $\Pi=2^L$ the partition size (number of summed coefficients by a measurement) in the initial tree level $L$, and
  \begin{equation}
    \label{eq:precondition_r}
    r=\sum_{n=k+1}^{2\Pi-1} |a_{h_n}| \,.
  \end{equation} 
  $K$-AHS will collect all significant coefficients $a_n\in  {\cal K}$, if
  \begin{equation}
    \label{eq:precondition_s}
    u>r \,.
  \end{equation}
\end{theorem}
\begin{IEEEproof}
  If this were not true, then there is a measurement containing significant coefficients, which is not larger than a measurement containing only non-significant coefficients. Let $\cal A$ be the set of coefficients of the measurement containing significant coefficients ($\cal A\cap \cal K\neq \emptyset$), and $\cal B$ be the set of coefficients of the measurement containing only non-significant coefficients ($\cal B\cap \cal K = \emptyset$). Then the following inequality
  \begin{eqnarray}
    \left\vert \sum_{a_n \in \cal A} a_n \right\vert &\leq &  \left\vert  \sum_{a_n \in  \cal B} a_n \right\vert\,
  \end{eqnarray}
  would hold. This can be written as
  \begin{eqnarray}
    \left\vert \sum_{a_n \in \cal A\cap \cal K} a_n + \sum_{a_n \in \cal A\setminus (\cal A\cap \cal K)} a_n\right\vert &\leq&  \left\vert  \sum_{a_n \in  \cal B} a_n \right\vert\,,
  \end{eqnarray}
  from which follows
  \begin{eqnarray}
    \left\vert \sum_{a_n \in \cal A\cap \cal K} a_n \right\vert &\leq&  \left\vert  \sum_{a_n \in  \cal B} a_n\right\vert + \left\vert  \sum_{a_n \in \cal A\setminus (\cal A\cap \cal K)} a_n\right\vert\, \\
    &\leq&  \sum_{a_n \in  \cal B} \left\vert  a_n\right\vert +  \sum_{a_n \in \cal A\setminus (\cal A\cap \cal K)} \left\vert a_n\right\vert \\
    &\leq&r\, .
  \end{eqnarray}
  Since $u$ is smaller or equal than the left hand side, this contradicts (\ref{eq:precondition_s}).
\end{IEEEproof}
With Theorem 2 we can analyze the $K$-AHS sensing quality for the introduced signal models.

\subsubsection{Application of Theorem \ref{th2} to the $k$-Sparse Model}
\label{sec:application_th2_k-sparse_model}

In the case of exactly $k$-sparse signals whose $k$ non-zero coefficients are drawn from a continuous probability distribution (signal model \ref{sec:l0-sparse_signal_model}), condition (\ref{eq:precondition_s}) 
holds almost surely for any $k\leq K$, since $r=0$ and $u>0$ with overwhelming probability.

\subsubsection{Application of Theorem \ref{th2} to the Exponential Model}
\label{sec:application_th2_exponential_model}

In the case of exponentially decaying coefficients (signal model \ref{sec:exp_decay_signal_model}), condition (\ref{eq:precondition_s}) holds for any $k\leq K$, if model parameter $q \geq 2$.
It can be easily seen that $u \geq Rq^{-k}$. For the right hand side of (\ref{eq:precondition_s}) we have
\begin{eqnarray}
  r = \sum_{n=k+1}^{2\Pi-1} |a_{h_n}| &<& \sum_{n=k+1}^{\infty} |a_{h_n}| = R q^{-k} \frac{1}{q-1}\\
  &<& R q^{-k} \leq u\,. \nonumber
\end{eqnarray}

\subsubsection{Application of Theorem \ref{th2} to the Power Law Model}
\label{sec:application_th2_power_model}

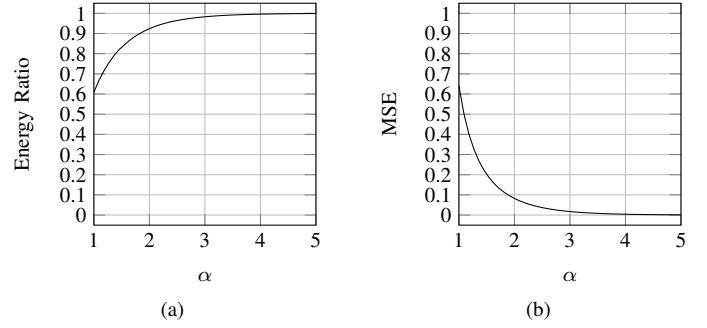
\begin{figure}[t!]
  \subfloat[]{
    \centering
    
  \begin{tikzpicture}
  \footnotesize
  \begin{axis}[
    width=.25\textwidth,
    height=.25\textwidth,
    xmin=1, xmax=5,
    ymin=-0.05, ymax=1.05,
    xtick={1,2,3,4,5,6,7,8,9,10},
    xticklabels={1,2,3,4,5,6,7,8,9,10},
    ytick={0.00, 0.10, 0.20, 0.30, 0.40, 0.50, 0.60, 0.70, 0.80, 0.90, 1.00},
    yticklabels={0, 0.1, 0.2, 0.3, 0.4, 0.5, 0.6, 0.7, 0.8, 0.9, 1},
    xlabel={$\alpha$},
    ylabel={Energy Ratio},
    ylabel style={yshift=-.3cm},
    xmajorgrids,
    ymajorgrids,
    ]

    \addplot [
    color=black,
    solid
    ]
    coordinates{ 
      (1.00,0.60937136) (1.09,0.66619487) (1.18,0.71434291) (1.27,0.75509219) (1.36,0.78961962) (1.45,0.81893713) (1.55,0.84389116) (1.64,0.86518295) (1.73,0.88339238) (1.82,0.89899980) (1.91,0.91240426) (2.00,0.92393842) (2.09,0.93388057) (2.18,0.94246431) (2.27,0.94988630) (2.36,0.95631268) (2.45,0.96188414) (2.55,0.96672015) (2.64,0.97092243) (2.73,0.97457776) (2.82,0.97776034) (2.91,0.98053376) (3.00,0.98295259) (3.09,0.98506378) (3.18,0.98690775) (3.27,0.98851938) (3.36,0.98992880) (3.45,0.99116209) (3.55,0.99224181) (3.64,0.99318756) (3.73,0.99401633) (3.82,0.99474291) (3.91,0.99538013) (4.00,0.99593920) (4.09,0.99642987) (4.18,0.99686063) (4.27,0.99723892) (4.36,0.99757122) (4.45,0.99786320) (4.55,0.99811980) (4.64,0.99834536) (4.73,0.99854368) (4.82,0.99871808) (4.91,0.99887147) (5.00,0.99900641) (5.09,0.99912514) (5.18,0.99922960) (5.27,0.99932154) (5.36,0.99940247) (5.45,0.99947370) (5.55,0.99953641) (5.64,0.99959163) (5.73,0.99964025) (5.82,0.99968306) (5.91,0.99972077) (6.00,0.99975397) (6.09,0.99978322) (6.18,0.99980899) (6.27,0.99983169) (6.36,0.99985168) (6.45,0.99986930) (6.55,0.99988482) (6.64,0.99989849) (6.73,0.99991054) (6.82,0.99992116) (6.91,0.99993051) (7.00,0.99993876) (7.09,0.99994602) (7.18,0.99995242) (7.27,0.99995807) (7.36,0.99996304) (7.45,0.99996742) (7.55,0.99997128) (7.64,0.99997469) (7.73,0.99997769) (7.82,0.99998033) (7.91,0.99998266) (8.00,0.99998472) (8.09,0.99998653) (8.18,0.99998813) (8.27,0.99998953) (8.36,0.99999077) (8.45,0.99999187) (8.55,0.99999283) (8.64,0.99999368) (8.73,0.99999443) (8.82,0.99999509) (8.91,0.99999567) (9.00,0.99999618) (9.09,0.99999663) (9.18,0.99999703) (9.27,0.99999738) (9.36,0.99999769) (9.45,0.99999797) (9.55,0.99999821) (9.64,0.99999842) (9.73,0.99999861) (9.82,0.99999877) (9.91,0.99999892) (10.00,0.99999905)
    };
  \end{axis}
\end{tikzpicture}

    \label{fig:rel_energy_1st_coeff}
  }
  \subfloat[]{
    \centering
    
  \begin{tikzpicture}
  \footnotesize
  \begin{axis}[
    width=.25\textwidth,
    height=.25\textwidth,
    xmin=1, xmax=5,
    ymin=-0.05, ymax=1.05,
    xtick={1,2,3,4,5,6,7,8,9,10},
    xticklabels={1,2,3,4,5,6,7,8,9,10},
    ytick={0.00, 0.10, 0.20, 0.30, 0.40, 0.50, 0.60, 0.70, 0.80, 0.90, 1.00},
    yticklabels={0, 0.1, 0.2, 0.3, 0.4, 0.5, 0.6, 0.7, 0.8, 0.9, 1},
    xlabel={$\alpha$},
    ylabel={MSE},
    ylabel style={yshift=-.3cm},
    xmajorgrids,
    ymajorgrids,
    ]

    \addplot [
    color=black,
    solid
    ]
    coordinates{
      (1.00,0.64103544) (1.09,0.50224198) (1.18,0.40162690) (1.27,0.32632023) (1.36,0.26848061) (1.45,0.22311996) (1.54,0.18693862) (1.63,0.15767675) (1.72,0.13373905) (1.81,0.11396934) (1.90,0.09751051) (1.99,0.08371455) (2.08,0.07208315) (2.17,0.06222743) (2.26,0.05383998) (2.35,0.04667500) (2.44,0.04053400) (2.53,0.03525527) (2.62,0.03070604) (2.71,0.02677655) (2.80,0.02337548) (2.89,0.02042643) (2.98,0.01786519) (3.07,0.01563753) (3.16,0.01369747) (3.25,0.01200590) (3.34,0.01052943) (3.43,0.00923947) (3.52,0.00811148) (3.61,0.00712435) (3.70,0.00625988) (3.79,0.00550232) (3.88,0.00483807) (3.97,0.00425531) (4.06,0.00374380) (4.15,0.00329463) (4.24,0.00290003) (4.33,0.00255325) (4.42,0.00224838) (4.51,0.00198027) (4.60,0.00174443) (4.69,0.00153692) (4.78,0.00135428) (4.87,0.00119350) (4.96,0.00105194) (5.05,0.00092727) (5.14,0.00081746) (5.23,0.00072073) (5.32,0.00063549) (5.41,0.00056038) (5.50,0.00049419) (5.59,0.00043584) (5.68,0.00038441) (5.77,0.00033907) (5.86,0.00029909) (5.95,0.00026384) (6.04,0.00023275) (6.13,0.00020534) (6.22,0.00018116) (6.31,0.00015983) (6.40,0.00014102) (6.49,0.00012443) (6.58,0.00010979) (6.67,0.00009688) (6.76,0.00008549) (6.85,0.00007544) (6.94,0.00006657) (7.03,0.00005875) (7.12,0.00005184) (7.21,0.00004575) (7.30,0.00004038) (7.39,0.00003563) (7.48,0.00003145) (7.57,0.00002776) (7.66,0.00002450) (7.75,0.00002162) (7.84,0.00001908) (7.93,0.00001684) (8.02,0.00001486) (8.11,0.00001312) (8.20,0.00001158) (8.29,0.00001022) (8.38,0.00000902) (8.47,0.00000796) (8.56,0.00000703) (8.65,0.00000620) (8.74,0.00000547) (8.83,0.00000483) (8.92,0.00000427) (9.01,0.00000376) (9.10,0.00000332) (9.19,0.00000293) (9.28,0.00000259) (9.37,0.00000229) (9.46,0.00000202) (9.55,0.00000178) (9.64,0.00000157) (9.73,0.00000139) (9.82,0.00000122) (9.91,0.00000108) (10.00,0.00000095)
    };
  \end{axis}
\end{tikzpicture}

    \label{fig:mse_approx_1st_coeff}
  }
  \caption{Relevance of the most significant coefficient $a_{h_1}$ for power law decaying coefficients. (a) Energy ratio between optimal $1$-term approximation of $\vec{x}$ and the full signal $\vec{x}$ depending on $\alpha$. (b) Mean squared error (MSE) between optimal $1$-term approximation of $\vec{x}$ and the full signal $\vec{x}$ depending on $\alpha$.}
\end{figure}

In the case of power law decaying coefficients (signal model \ref{sec:power_law_decay_signal_model}), Theorem \ref{th2} cannot be applied directly for all cases in which $k \leq K$. It allows, nevertheless, to state conditions on model parameter $\alpha$ for the case $k=1$, meaning that the detection of $a_{h_1}$, the most significant coefficient, is guaranteed. This is useful since the bulk of the signal energy often lies in the first coefficient. For this model, \figurename\ \ref{fig:rel_energy_1st_coeff} illustrates the energy ratio between optimal $1$-term approximation of the signal and the complete signal as a function of model parameter $\alpha$. An increase of $\alpha$ rapidly concentrates the signal energy on $a_{h_1}$ such that this coefficient contributes nearly exclusively to the entire energy of the signal. A similar illustration is provided by \figurename\ \ref{fig:mse_approx_1st_coeff} in terms of mean squared error (MSE). 

For $k=1$, condition (\ref{eq:precondition_s}) of Theorem \ref{th2} holds, if $\alpha>\alpha^*$ with $\alpha^*$ being defined by
\begin{eqnarray}
  \label{eq:alpha_star}
  \sum_{n=2}^{\infty} n^{-\alpha^*} =\zeta(\alpha^*)-1 =1\,,
\end{eqnarray}
where $\zeta(\cdot)$ denotes the Riemann zeta function. The value of $\alpha^*$ is about $1.73$. Since $k=1$, we have $u=R$ and furthermore $r<R$, due to (\ref{eq:alpha_star}). 
Hence, if $\alpha>\alpha^*$, we can guarantee that $K$-AHS captures more than 88\% of the signal energy.

\begin{figure}[t!]
  \centering
  
  \input{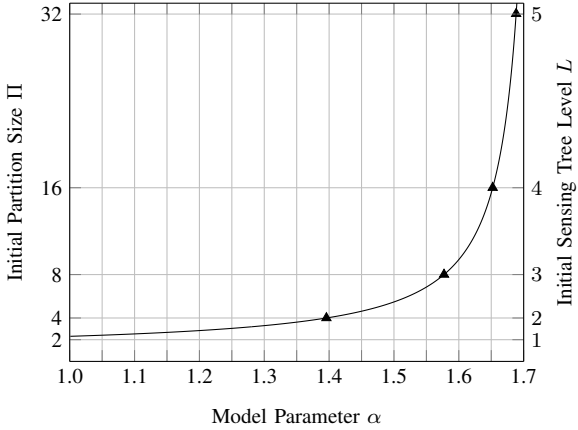}

  \caption{Initial partition size $\Pi$ (left vertical axis) and the corresponding initial sensing tree level $L$ (right vertical axis) dependent on model parameter $\alpha$ as stated by (\ref{eq:upper_bound_partition_size}) for signals with power law decaying coefficients (see Section \ref{sec:power_law_decay_signal_model}). For a particular value of $\alpha$, any power to base 2 under the curve is a valid initial partition size in order to capture the most prominent signal coefficient $a_{h_1}$ by $K$-AHS. The vertical axis on the right indicates for each $\Pi$ the corresponding $L$. The triangular markers indicate values of alpha for which the integer valued partition size $\Pi$ has to be modified.}
  \label{fig:upper_bound_partition_size}
\end{figure}

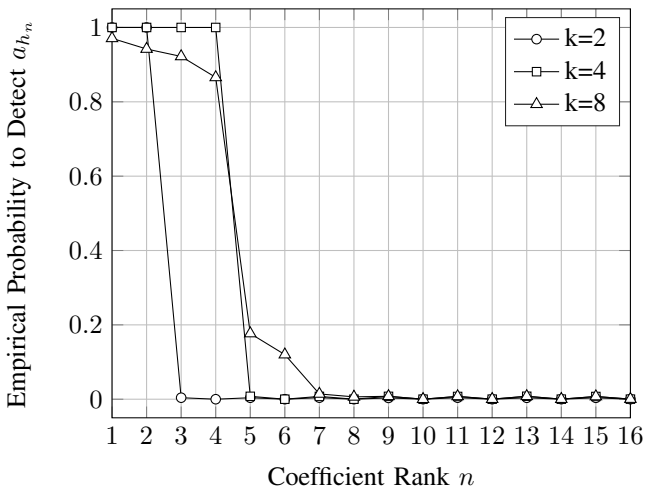
\begin{figure}[b!]
  \centering
  
%
%
\begin{tikzpicture}

\begin{axis}[%
width=0.38\textwidth,
height=0.3\textwidth,
at={(0\textwidth,0\textwidth)},
scale only axis,
xmin=1,
xmax=16,
xtick={ 1,  2,  3,  4,  5,  6,  7,  8,  9, 10, 11, 12, 13, 14, 15, 16},
xlabel={Coefficient Rank $n$},
xmajorgrids,
ymin=-0.05,
ymax=1.05,
ylabel={Empirical Probability to Detect $a_{h_n}$},
ymajorgrids,
axis background/.style={fill=white},
legend style={legend cell align=left,align=left,draw=white!15!black}
]
\addplot [color=black,solid,mark size=1.8pt,mark=*,mark options={solid,fill=white}]
  table[row sep=crcr]{%
1	0.999999999998084\\
2	0.999999999998084\\
3	0.00406\\
4	0\\
5	0.00398000000000001\\
6	1e-05\\
7	0.00417\\
8	3e-05\\
9	0.00348000000000001\\
10	6e-05\\
11	0.00409\\
12	2e-05\\
13	0.00384000000000001\\
14	9e-05\\
15	0.00377000000000001\\
16	4e-05\\
};
\addlegendentry{k=2};

\addplot [color=black,solid,mark size=1.6pt,mark=square*,mark options={solid,fill=white}]
  table[row sep=crcr]{%
1	0.999999999998084\\
2	0.999999999998084\\
3	0.999999999998084\\
4	0.999999999998084\\
5	0.00774999999999985\\
6	6e-05\\
7	0.00755999999999986\\
8	8e-05\\
9	0.00765999999999986\\
10	0.00016\\
11	0.00738999999999987\\
12	0.0002\\
13	0.00791999999999985\\
14	0.00017\\
15	0.00732999999999987\\
16	0.00023\\
};
\addlegendentry{k=4};

\addplot [color=black,solid,mark size=2.5pt,mark=triangle*,mark options={solid,fill=white}]
  table[row sep=crcr]{%
1	0.970719999998217\\
2	0.942059999998347\\
3	0.922059999998438\\
4	0.865689999998695\\
5	0.176720000000036\\
6	0.119969999999986\\
7	0.0141999999999996\\
8	0.00633999999999991\\
9	0.00787999999999985\\
10	4e-05\\
11	0.00767999999999986\\
12	0.00016\\
13	0.00785999999999985\\
14	0.00021\\
15	0.00729999999999987\\
16	0.00036\\
};
\addlegendentry{k=8};

\end{axis}
\end{tikzpicture}%

  \caption{$K$-AHS simulation with randomly generated synthetic $k$-sparse signals (see Section \ref{sec:l0-sparse_signal_model}), locations uniformly distributed, non-zero coefficients standard Gaussian distributed. For model parameter $k \in \{2, 4, 8\}$, $10^5$ signals of dimensionality $N=1024$ were generated. The empirical $K$-AHS detection probability for the $16$ most significant coefficients is plotted according to their rank. $K$-AHS was applied with user parameter $K=4$. As long as $K \geq k$, all $k$ non-zero coefficients are identified correctly.}
  \label{fig:empirical_probability_detect_coef_l0-model}
\end{figure}

Note that this finding does not depend on the initial partition size $\Pi$. By considering $\Pi$, the detection of $a_{h_1}$ can be guaranteed for even smaller values of $\alpha$. By using integral approximations of the partial sum (\ref{eq:precondition_r}), we obtain
\begin{eqnarray}
  \nonumber
  r &=&  \sum_{n=2}^{2\Pi-1} n^{-\alpha} \\
  \nonumber
  &\leq& 2^{-\alpha} + \int\limits_{\frac{5}{2}}^{2\Pi-\frac{1}{2}} x^{-\alpha} \mathrm{d}x\\
  &\leq& 2^{-\alpha} + \frac{1}{1-\alpha} \left( \left(2\Pi-\frac{1}{2} \right)^{1-\alpha} - \left(\frac{5}{2}\right)^{1-\alpha} \right)
  \label{eq:upper_bound_partition_size}
\end{eqnarray}

If we choose $K$ such that we start with a value $\Pi$ for which the r.h.s.\ of (\ref{eq:upper_bound_partition_size}) is smaller than $1$, then the most significant coefficient $a_{h_1}$ is definitely captured by $K$-AHS. \figurename\ \ref{fig:upper_bound_partition_size} plots the maximal $\Pi$, which is allowed due to (\ref{eq:upper_bound_partition_size}), as a function of $\alpha$.

\section{Results}
\label{sec:results}

\subsection{Synthetic Signals}

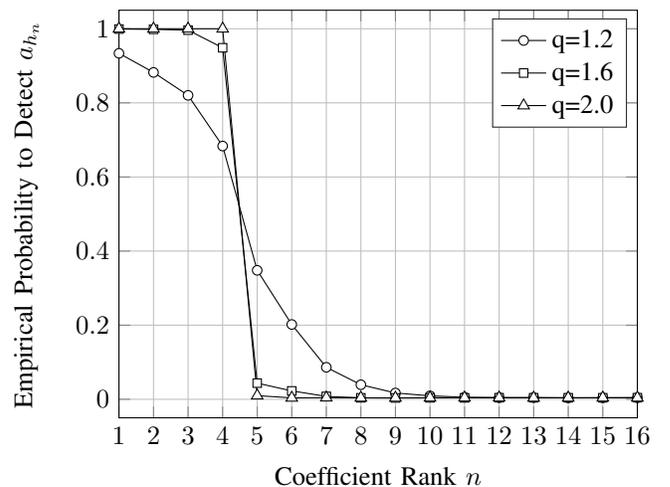
\begin{figure}[b!]
  \centering
  
%
%
\begin{tikzpicture}

\begin{axis}[%
width=0.38\textwidth,
height=0.299\textwidth,
at={(0\textwidth,0\textwidth)},
scale only axis,
xmin=1,
xmax=16,
xtick={ 1,  2,  3,  4,  5,  6,  7,  8,  9, 10, 11, 12, 13, 14, 15, 16},
xlabel={Coefficient Rank $n$},
xmajorgrids,
ymin=-0.05,
ymax=1.05,
ylabel={Empirical Probability to Detect $a_{h_n}$},
ymajorgrids,
axis background/.style={fill=white},
legend style={legend cell align=left,align=left,draw=white!15!black}
]
\addplot [color=black,solid,mark size=1.8pt,mark=*,mark options={solid,fill=white}]
  table[row sep=crcr]{%
1	0.933819999998385\\
2	0.88225999999862\\
3	0.820269999998902\\
4	0.683539999999524\\
5	0.348030000000207\\
6	0.201860000000061\\
7	0.0862799999999992\\
8	0.0392500000000013\\
9	0.0169799999999995\\
10	0.00930999999999979\\
11	0.00554999999999994\\
12	0.00422\\
13	0.00457999999999998\\
14	0.00376000000000001\\
15	0.00399000000000001\\
16	0.00405\\
};
\addlegendentry{q=1.2};

\addplot [color=black,solid,mark size=1.6pt,mark=square*,mark options={solid,fill=white}]
  table[row sep=crcr]{%
1	0.999249999998087\\
2	0.998429999998091\\
3	0.996309999998101\\
4	0.948679999998317\\
5	0.0435700000000027\\
6	0.0225799999999992\\
7	0.00777999999999985\\
8	0.00395000000000001\\
9	0.00361000000000001\\
10	0.00386000000000001\\
11	0.00383000000000001\\
12	0.00422\\
13	0.00411\\
14	0.00415\\
15	0.00405\\
16	0.00429999999999999\\
};
\addlegendentry{q=1.6};

\addplot [color=black,solid,mark size=2.5pt,mark=triangle*,mark options={solid,fill=white}]
  table[row sep=crcr]{%
1	0.999999999998084\\
2	0.999999999998084\\
3	0.999999999998084\\
4	0.999999999998084\\
5	0.00974999999999977\\
6	0.00382000000000001\\
7	0.00426999999999999\\
8	0.00382000000000001\\
9	0.00375000000000001\\
10	0.00406\\
11	0.00417\\
12	0.00419\\
13	0.00384000000000001\\
14	0.00394000000000001\\
15	0.00408\\
16	0.00404\\
};
\addlegendentry{q=2.0};

\end{axis}
\end{tikzpicture}%

  \caption{$K$-AHS simulation with randomly generated synthetic signals with exponentially decaying coefficients (see Section \ref{sec:exp_decay_signal_model}), locations and signs uniformly distributed. For model parameter $q \in \{1.2, 1.6, 2\}$, $10^5$ signals of dimensionality $N=1024$ were generated. The empirical $K$-AHS detection probability for the $16$ most significant coefficients is plotted according to their rank. $K$-AHS was applied with user parameter $K=4$. As long as $q \geq 2$, the $K$ most significant coefficients are identified correctly.}
  \label{fig:empirical_probability_detect_coef_exp-model}
\end{figure}

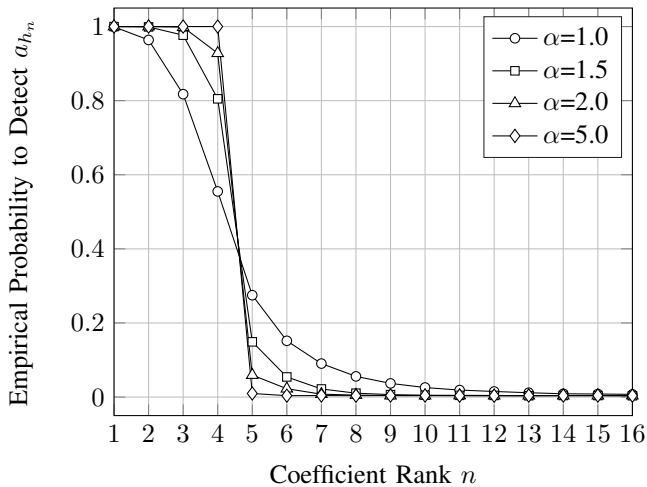
\begin{figure}[b!]
  \centering
  
%
%
\begin{tikzpicture}

\begin{axis}[%
width=0.38\textwidth,
height=0.299\textwidth,
scale only axis,
xmin=1,
xmax=16,
xtick={ 1,  2,  3,  4,  5,  6,  7,  8,  9, 10, 11, 12, 13, 14, 15, 16},
xlabel={Coefficient Rank $n$},
xmajorgrids,
ymin=-0.05,
ymax=1.05,
ylabel={Empirical Probability to Detect  $a_{h_n}$},
ymajorgrids,
axis background/.style={fill=white},
legend style={legend cell align=left,align=left,draw=white!15!black}
]
\addplot [color=black,solid,mark size=1.8pt,mark=*,mark options={solid,fill=white}]
  table[row sep=crcr]{%
1	0.999199999998087\\
2	0.963829999998248\\
3	0.817549999998914\\
4	0.555020000000109\\
5	0.274980000000134\\
6	0.151830000000011\\
7	0.0903899999999977\\
8	0.0558800000000064\\
9	0.0369100000000006\\
10	0.0257199999999991\\
11	0.0189099999999994\\
12	0.0150999999999996\\
13	0.0116899999999997\\
14	0.00887999999999981\\
15	0.00819999999999983\\
16	0.00729999999999987\\
};
\addlegendentry{$\alpha\text{=1.0}$};

\addplot [color=black,solid,mark size=1.6pt,mark=square*,mark options={solid,fill=white}]
  table[row sep=crcr]{%
1	0.999999999998084\\
2	0.999039999998088\\
3	0.97671999999819\\
4	0.804989999998971\\
5	0.149070000000008\\
6	0.054300000000006\\
7	0.0218699999999993\\
8	0.0105299999999997\\
9	0.0066199999999999\\
10	0.00495999999999997\\
11	0.00463999999999998\\
12	0.00442999999999999\\
13	0.00352000000000001\\
14	0.0042\\
15	0.00403\\
16	0.00359000000000001\\
};
\addlegendentry{$\alpha\text{=1.5}$};

\addplot [color=black,solid,mark size=2.5pt,mark=triangle*,mark options={solid,fill=white}]
  table[row sep=crcr]{%
1	0.999999999998084\\
2	0.999999999998084\\
3	0.998449999998091\\
4	0.927949999998412\\
5	0.0593300000000075\\
6	0.0225299999999993\\
7	0.00783999999999985\\
8	0.00530999999999995\\
9	0.00390000000000001\\
10	0.00385000000000001\\
11	0.00423\\
12	0.00378000000000001\\
13	0.00386000000000001\\
14	0.00374000000000001\\
15	0.00379000000000001\\
16	0.00392000000000001\\
};
\addlegendentry{$\alpha\text{=2.0}$};

\addplot [color=black,solid,mark size=2.5pt,mark=diamond*,mark options={solid,fill=white}]
  table[row sep=crcr]{%
1	0.999999999998084\\
2	0.999999999998084\\
3	0.999999999998084\\
4	0.999999999998084\\
5	0.00967999999999977\\
6	0.00378000000000001\\
7	0.00409\\
8	0.00375000000000001\\
9	0.00413\\
10	0.00421\\
11	0.00368000000000001\\
12	0.00362000000000001\\
13	0.00405\\
14	0.00406\\
15	0.0041\\
16	0.00376000000000001\\
};
\addlegendentry{$\alpha\text{=5.0}$};

\end{axis}
\end{tikzpicture}%

  \caption{$K$-AHS simulation with randomly generated synthetic signals with power law decaying coefficients (see Section \ref{sec:power_law_decay_signal_model}), locations and signs uniformly distributed. For model parameter $\alpha \in \{1, 1.5, 2, 5\}$, $10^5$ signals of dimensionality $N=1024$ were generated. The empirical $K$-AHS detection probability for the $16$ most significant coefficients is plotted according to their rank. $K$-AHS was applied with user parameter $K=4$. The most significant coefficient $a_{h_1}$ is nearly always found, even for $\alpha < \alpha^*$.}
  \label{fig:empirical_probability_detect_coef_power-model}
\end{figure}

We conducted sensing experiments with $K$-AHS on synthetic signals that obey the models introduced in Section \ref{subsec:signal_models}. To complement our theoretical findings of Section \ref{sec:SSCI}, we empirically study the performance of $K$-AHS to detect significant coefficients depending on the model parameters. For each parameter value we generated $10^5$ signals of dimensionality $N=1024$. First, the magnitudes of coefficients were computed as given by the model. Second, locations and signs of the coefficients were assigned uniformly at random. Subsequently, we applied $K$-AHS to each signal by setting the user parameter to $K=4$, and calculated the empirical detection probability for individual coefficient ranks. Ideally, the empirical probability for each coefficient $a_{h_1}, ..., a_{h_K}$ is equal or close to $1$. \figurename s \ref{fig:empirical_probability_detect_coef_l0-model} to \ref{fig:empirical_probability_detect_coef_power-model} show this empirical detection probability 
for the three different signal models and the $16$ most significant ranks (out of $1024$).

\figurename\ \ref{fig:empirical_probability_detect_coef_l0-model} illustrates simulation results for the $k$-sparse signal model of Section \ref{sec:l0-sparse_signal_model}. While the number of non-zero coefficients is given by model parameter $k$, their values were drawn from a standard Gaussian distribution. The values of the model parameter that we investigated were $k \in \{2,4,8\}$. In the cases $k=2$ and $k=4$ all non-zero coefficients were identified correctly. This is in accordance with our theoretical finding in \ref{sec:application_th2_k-sparse_model} which predicts perfect recovery if $K \geq k$. In the case $k=8$, we have the situation $K < k$ and the empirical detection probability is decreased. However, it is still above $0.8$ for each $a_{h_1},...,a_{h_K}$ despite the fact that the number of non-zero coefficients of the signal is considerably underestimated.

\figurename\ \ref{fig:empirical_probability_detect_coef_exp-model} illustrates simulation results for the exponential decay model of Section \ref{sec:exp_decay_signal_model}.
The values of the model parameter that we investigated were $q \in \{1.2,1.6,2\}$. It can be seen that an increase of base $q$ (steeper decay of coefficients) leads to an increase of the empirical detection probability for $a_{h_1},...,a_{h_K}$. All of the $K$ most prominent coefficients are identified correctly in the scenario $q = 2$, which is predicted by our theoretical finding in \ref{sec:application_th2_exponential_model}.

\begin{figure}[b!]
  \centering
  
%
%
\begin{tikzpicture}

\begin{axis}[%
width=0.38\textwidth,
height=0.299\textwidth,
scale only axis,
xmin=1,
xmax=32,
xtick={ 2,  4,  8, 12, 16, 20, 24, 28, 32},
xlabel={User Parameter $K$},
xmajorgrids,
ymin=0.45,
ymax=1.05,
ytick={.5, .55, .6, .65, .7, .75, .8, .85, .9, .95, 1},
yticklabels={0.5, , 0.6, , 0.7, , 0.8, , 0.9, , 1},
ylabel={Relative Signal Energy},
ymajorgrids,
axis background/.style={fill=white},
legend style={at={(0.97,0.03)},anchor=south east,legend cell align=left,align=left,draw=white!15!black}
]
\addplot [color=black,solid,mark size=1.8pt,mark=*,mark options={solid,fill=white}]
 plot [error bars/.cd, y dir = both, y explicit]
 table[row sep=crcr, y error plus index=2, y error minus index=3]{%
2	0.841602153834444	0.0461032136516026	0.0461032136516026\\
4	0.928492916937599	0.019762927978789	0.019762927978789\\
8	0.970291428379967	0.0081650338949778	0.0081650338949778\\
12	0.98203371360471	0.00600449797939939	0.00600449797939939\\
16	0.988294719754583	0.00299683807125513	0.00299683807125513\\
20	0.991072768950319	0.00306247158063977	0.00306247158063977\\
24	0.992724379795194	0.00306080700564367	0.00306080700564367\\
28	0.994068478256501	0.00259169937648618	0.00259169937648618\\
32	0.9954434015509	0.00140468050446245	0.00140468050446245\\
};
\addlegendentry{$\alpha\text{=1.2}$};

\addplot [color=black,solid,mark size=1.6pt,mark=square*,mark options={solid,fill=white}]
 plot [error bars/.cd, y dir = both, y explicit]
 table[row sep=crcr, y error plus index=2, y error minus index=3]{%
2	0.931167159104603	0.0193525286978106	0.0193525286978106\\
4	0.97775487456789	0.00538307934627925	0.00538307934627925\\
8	0.993425570476624	0.00315556687242432	0.00315556687242432\\
12	0.99687886019776	0.00186353104394751	0.00186353104394751\\
16	0.998268919754713	0.000674994452179521	0.000674994452179521\\
20	0.998834456382605	0.000489834065065651	0.000489834065065651\\
24	0.999172187265706	0.000386797941749049	0.000386797941749049\\
28	0.999373152186637	0.00042627824029748	0.00042627824029748\\
32	0.999555317326874	0.000154934224556438	0.000154934224556438\\
};
\addlegendentry{$\alpha\text{=1.5}$};

\addplot [color=black,solid,mark size=2.5pt,mark=triangle*,mark options={solid,fill=white}]
 plot [error bars/.cd, y dir = both, y explicit]
 table[row sep=crcr, y error plus index=2, y error minus index=3]{%
2	0.981728874368217	0.000562781144574315	0.000562781144574315\\
4	0.996467275915512	0.00127106206080016	0.00127106206080016\\
8	0.999462085064126	0.000223656406888829	0.000223656406888829\\
12	0.999822574892558	6.74832749766321e-05	6.74832749766321e-05\\
16	0.99992387643588	3.78879294349597e-05	3.78879294349597e-05\\
20	0.999961037945123	1.3702735504582e-05	1.3702735504582e-05\\
24	0.999975765562801	1.37998353506092e-05	1.37998353506092e-05\\
28	0.99998406130365	1.21074990116954e-05	1.21074990116954e-05\\
32	0.999989858795574	9.12964486389402e-06	9.12964486389402e-06\\
};
\addlegendentry{$\alpha\text{=2.0}$};

\addplot [color=black,solid,mark size=2.5pt,mark=diamond*,mark options={solid,fill=white}]
 plot [error bars/.cd, y dir = both, y explicit]
 table[row sep=crcr, y error plus index=2, y error minus index=3]{%
2	0.999982058056884	9.2659640198939e-07	9.2659640198939e-07\\
4	0.999999877804989	1.22225557551106e-08	1.22225557551106e-08\\
8	0.999999999557232	6.4569188253691e-11	6.4569188253691e-11\\
12	0.99999999998581	2.62449814092642e-12	2.62449814092642e-12\\
16	0.999999999998834	2.19438877889217e-13	2.19438877889217e-13\\
20	0.999999999999823	9.14350940522029e-14	9.14350940522029e-14\\
24	0.99999999999998	1.68666578912706e-14	1.68666578912706e-14\\
28	0.999999999999998	1.96970602704679e-14	1.96970602704679e-14\\
32	1	3.0570918282866e-15	3.0570918282866e-15\\
};
\addlegendentry{$\alpha\text{=5.0}$};

\end{axis}
\end{tikzpicture}%

  \caption{$K$-AHS simulation with randomly generated synthetic signals with power law decaying coefficients (see Section \ref{sec:power_law_decay_signal_model}), locations and signs uniformly distributed. For model parameter $\alpha \in \{1.2, 1.5, 2, 5\}$, $10^3$ signals of dimensionality $N=1024$ were generated. The relative signal energy obtained by $K$-AHS is plotted dependent on user parameter $K$.\newline \textcolor{white}{q}\newline \textcolor{white}{q}}
  \label{fig:power_law_relative_signal_energy_vs_K}
\end{figure}
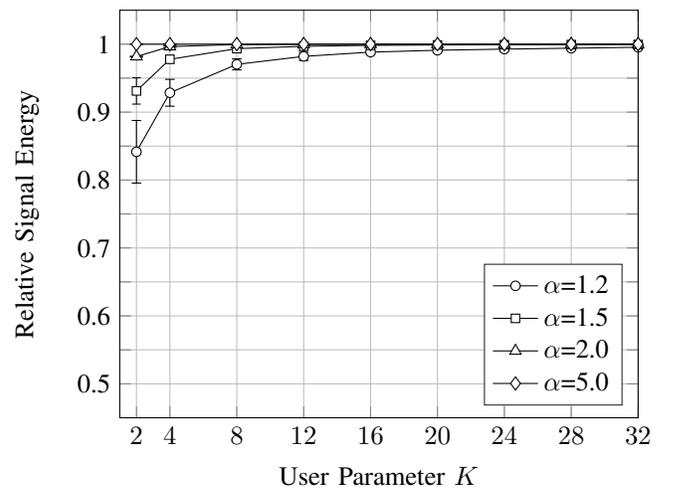

\figurename\ \ref{fig:empirical_probability_detect_coef_power-model} illustrates simulation results for the power decay model of Section \ref{sec:power_law_decay_signal_model}. As for the exponential model, a larger parameter value $\alpha$ results in a steeper decay of coefficients and increases the detection probability for significant coefficients. As opposed to the exponential model, a single threshold of model parameter $\alpha$ does not guarantee the detection of the most prominent coefficients for all values of $K$. On the other hand, the signal energy rapidly focuses on $a_{h_1}$ as $\alpha$ increases, see \figurename\ \ref{fig:rel_energy_1st_coeff}. Therefore, we additionally illustrate for the power decay model the relative signal energy obtained by $K$-AHS dependent on $K$. \figurename\ \ref{fig:power_law_relative_signal_energy_vs_K} shows that, for various values of $\alpha$, the reconstruction performance in terms of captured signal energy increases as $K$ is set to higher values.

\subsection{Natural Images}

\begin{figure}[b!]
  \centering
  \subfloat[]{\includegraphics[width=.24\textwidth]{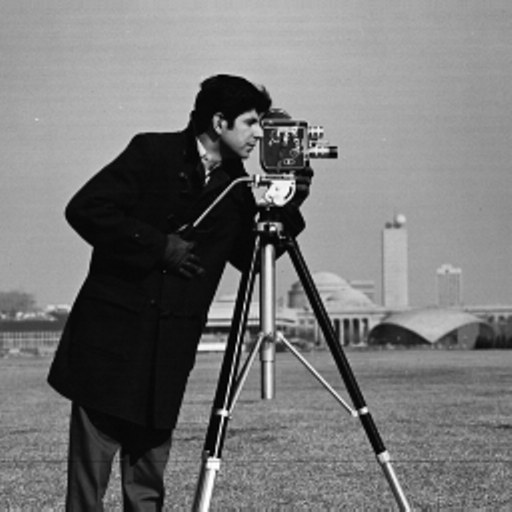}}
  \hfill
  \subfloat[]{\includegraphics[width=.24\textwidth]{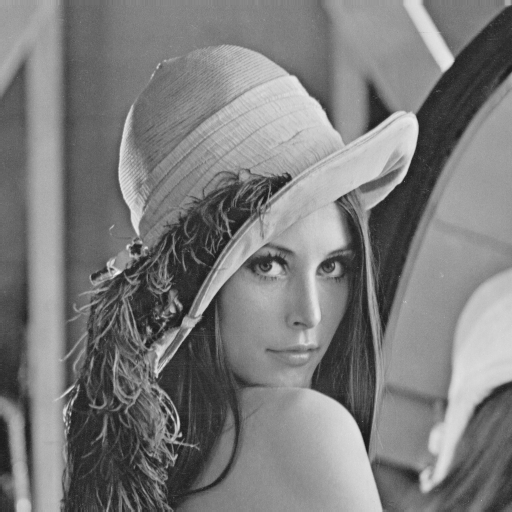}}
  
  \subfloat[]{\includegraphics[width=.24\textwidth]{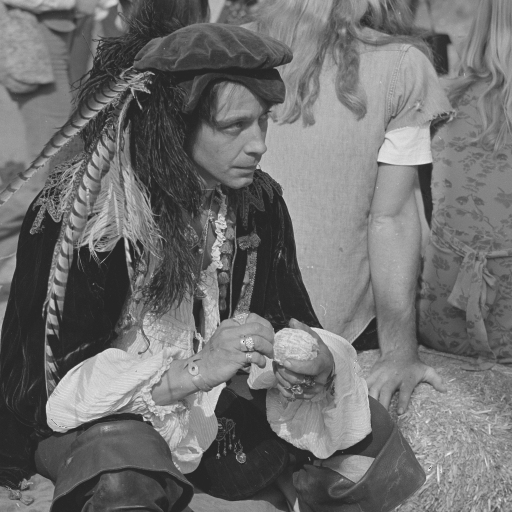}}
  \caption{Original test images with a size of $512\times512$ pixels and a gray level depth of 8 bit which were used for the compressive imaging experiments. (a) Test image \textit{Lena}. (b) Test image \textit{Cameraman}. (c) Test image \textit{Pirate}.}
  \label{fig:original_test_images}
\end{figure}

We conducted compressive imaging experiments by applying $K$-AHS to standard gray scale test images (\textit{Cameraman}, \textit{Lena}, \textit{Pirate}) with a size of $512\times512$ pixels and a gray level depth of 8 bit (see \figurename\ \ref{fig:original_test_images}). For each image, the reconstruction performance by $K$-AHS is measured in terms of peak-signal-to-noise ratio (PSNR). By use of (\ref{eq:exact_number_AHS_measurements}), the user parameter $K$ was varied such that the number of measurements $M$ took values from $0.02N$ to $0.3N$ in steps of $0.02N$. We report mean and standard deviation of PSNR over 10 trials. For each trial the sequence of basis vectors $\vec{\psi}_n$ was randomly shuffled before the assignment to the leaf nodes of the sensing tree. As sparse coding transform $\mat{\Psi}$ we chose (\textit{i}) the orthogonal non-standard 2D Haar wavelet basis, and (\textit{ii}) the biorthogonal Cohen-Daubechies-Feauveau 9/7 (CDF97) wavelet basis, which is part of the JPEG 2000 standard \cite{TaubmanMarcellin2001}.

\begin{figure}[b!]
  \centering
  \subfloat[]{
  \begin{tikzpicture}
  \footnotesize

  \begin{axis}[
    width=.5\textwidth,
    height=.22\textheight,
    xmin=.015, xmax=.305,
    ymin=17, ymax=35,
    axis x line*=top,
    hide y axis,
    xtick={.02,.04,.06,.08,.10,.12,.14,.16,.18,.20,.22,.24,.26,.28,.30},
    xticklabel style={rotate=90},
    xticklabels={0.0010,0.0023,0.0037,0.0054,0.0070,0.0089,0.0109,0.0129,0.0149,0.0172,0.0197,0.0222,0.0247,0.0272,0.0297},
    xlabel={$K/N$ ($K$-AHS parameter)},
    xlabel near ticks,
    ]
  \end{axis}
  
  \begin{axis}[
    width=.5\textwidth,
    height=.22\textheight,
    xmin=.015, xmax=.305,
    ymin=17, ymax=35,
    xtick={.02,.04,.06,.08,.10,.12,.14,.16,.18,.20,.22,.24,.26,.28,.30},
    xticklabel style={rotate=90},
    xticklabels={0.02,0.04,0.06,0.08,0.10,0.12,0.14,0.16,0.18,0.20,0.22,0.24,0.26,0.28,0.30},
    ytick={10,15,20,25,30,35,40,45,50,55,60},
    yticklabels={10,15,20,25,30,35,40,45,50,55,60},
    xlabel={$M/N$},
    xlabel near ticks,
    ylabel={PSNR},
    ylabel near ticks,
    xmajorgrids,
    ymajorgrids,
    legend entries={K-AHS (Haar), K-AHS (CDF97), CS (Haar), CS (CDF97)},
    legend style={font=\scriptsize,at={(0.45,1.4)},
      anchor=south,legend columns=-1}]

  \addplot [
  color=blue,
  solid
  ]
  plot [error bars/.cd, y dir = both, y explicit]
  coordinates{
    (0.0200,19.8372) +- (0,0.1125) (0.0400,21.5522) +- (0,0.1295) (0.0600,22.6755) +- (0,0.0877) (0.0800,23.7707) +- (0,0.1554) (0.1000,24.4826) +- (0,0.1415) (0.1200,25.4479) +- (0,0.1084) (0.1400,26.0633) +- (0,0.0909) (0.1600,26.5092) +- (0,0.1958) (0.1800,26.9173) +- (0,0.0989) (0.2000,27.7999) +- (0,0.0967) (0.2200,28.2784) +- (0,0.0631) (0.2400,28.6849) +- (0,0.0565) (0.2600,29.0541) +- (0,0.1118) (0.2800,29.4152) +- (0,0.1143) (0.3000,29.7114) +- (0,0.0887) 
  };

  \addplot [
  color=black,
  solid
  ]
  plot [error bars/.cd, y dir = both, y explicit]
  coordinates{
    (0.0200,20.4674) +- (0,0.1973) (0.0400,22.6388) +- (0,0.1399) (0.0600,24.0287) +- (0,0.1784) (0.0800,25.6144) +- (0,0.1291) (0.1000,26.4345) +- (0,0.0953) (0.1200,27.6345) +- (0,0.1698) (0.1400,28.3404) +- (0,0.2033) (0.1600,29.0687) +- (0,0.2048) (0.1800,29.5314) +- (0,0.2102) (0.2000,30.7612) +- (0,0.1122) (0.2200,31.3617) +- (0,0.2198) (0.2400,31.7779) +- (0,0.2713) (0.2600,32.3439) +- (0,0.2593) (0.2800,32.8507) +- (0,0.1374) (0.3000,33.3926) +- (0,0.1244) 
  };

  \addplot [
  color=blue,
  dashed
  ]
  plot [error bars/.cd, y dir = both, y explicit]
  coordinates{
    (0.0200,18.4566) (0.0400,20.0159) (0.0600,21.4021) (0.0800,22.6450) (0.1000,23.6571) (0.1200,24.5393) (0.1400,25.2858) (0.1600,26.0166) (0.1800,26.7838) (0.2000,27.2738) (0.2200,28.0813) (0.2400,28.7799) (0.2600,29.3604) (0.2800,30.0329) (0.3000,30.6476)
  };
  
  \addplot [
  color=black,
  dashed
  ]
  plot [error bars/.cd, y dir = both, y explicit]
  coordinates{
    (0.0200,18.0738) (0.0400,20.1508) (0.0600,21.9410) (0.0800,23.4799) (0.1000,24.7665) (0.1200,25.9785) (0.1400,26.9569) (0.1600,27.9222) (0.1800,28.9194) (0.2000,29.6939) (0.2200,30.5881) (0.2400,31.4298) (0.2600,32.2656) (0.2800,33.1263) (0.3000,33.8648) 
  };

\end{axis}
\end{tikzpicture}
}

  \subfloat[]{\includegraphics[width=.24\textwidth]{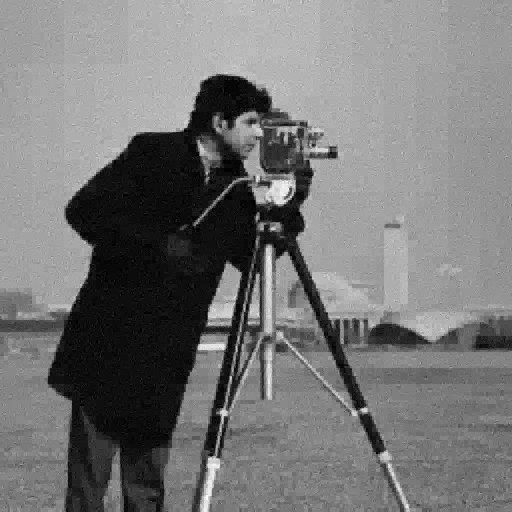}}
  \hfill
  \subfloat[]{\includegraphics[width=.24\textwidth]{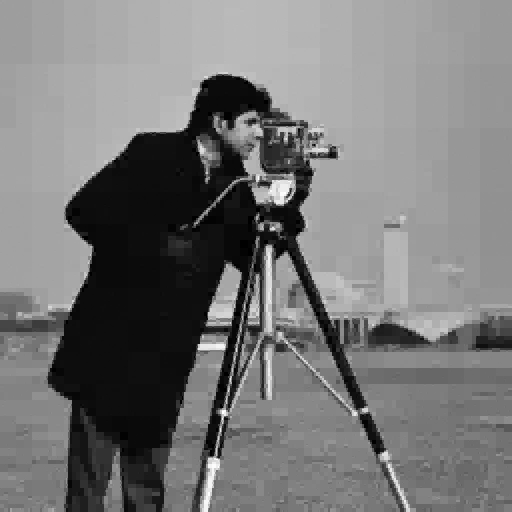}}

  \subfloat[]{\includegraphics[width=.24\textwidth]{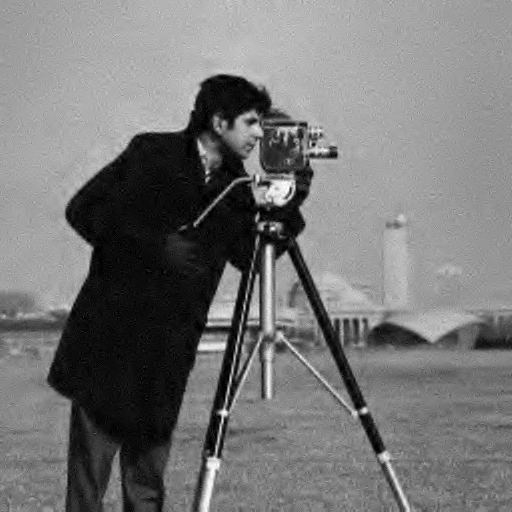}}
  \hfill
  \subfloat[]{\includegraphics[width=.24\textwidth]{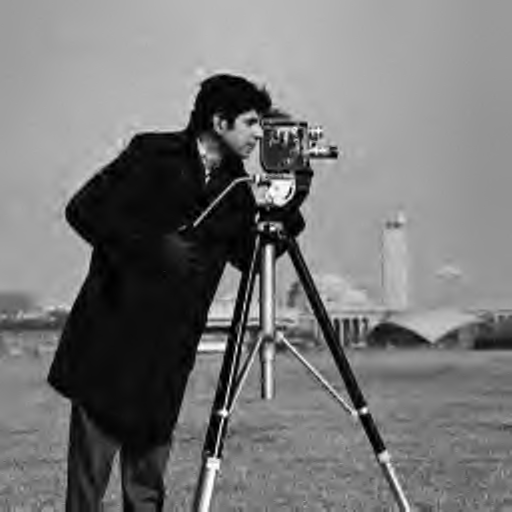}}

  \caption{Natural image sensing performance comparison between $K$-AHS and $\ell_1$-based CS for test image \textit{Cameraman}. (a) The PSNR dependent on the relative number of measurements. (b) CS reconstruction from $M=0.2N$ random noiselet measurements, Haar basis, PSNR: 27.27. (c) $K$-AHS reconstruction from $M=0.2N$ adaptive measurements, Haar basis, PSNR: 27.86. (d) CS reconstruction from $M=0.2N$ random noiselet measurements, CDF97 basis, PSNR: 29.69. (e) $K$-AHS reconstruction from $M=0.2N$ adaptive measurements, CDF97 basis, PSNR: 30.85. For visualization, reconstructed images were clipped to [0, 255] where necessary.}
  \label{fig:cameraman_results}
\end{figure}
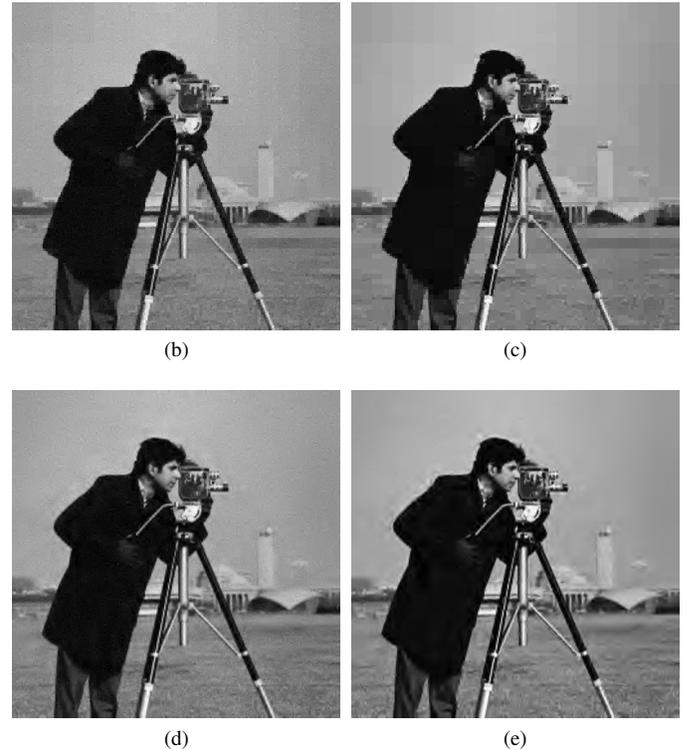

To provide a baseline comparison, we applied classical $\ell_1$-based CS to the test images as well. For the same sparse coding transforms, and the same values for $M$, linear measurements of each image were collected by sensing vectors that were randomly generated (without replacement) from the real valued noiselet transform. The random noiselet measurement ensemble was chosen in favor of CS due to its low coherence to the Haar basis \cite{TuHu09} and to the CDF97 basis \cite{PeLoSiCa14}. A low coherence between measurement ensemble $\mat{\Phi}$ and sparse transform $\mat{\Psi}$ assures that $\ell_1$-norm minimization recovers the original signal accurately \cite{CandesRomberg07}. In our classical CS experiments we addressed the following optimization problem
\begin{equation}
  \vec{a}^* = \argmin_{\vec{a} \in \mathbb{R}^N} \norm{\vec{a}}{1}\,\text{, s.t. } \mat{\Phi}\overline{\mat{\Psi}}^T\vec{a} = \vec{y}\,.
  \label{eq:l1_eq_optim_prob}
\end{equation}
In order to solve (\ref{eq:l1_eq_optim_prob}), we used the NESTA package \cite{BeBoCa11}, an $\ell_1$-recovery toolbox suited for solving large-scale compressed sensing reconstruction problems. NESTA is a cutting-edge first-order optimization procedure that exploits ideas from Nesterov \cite{Nesterov2005} such as accelerated descent methods and smoothing techniques.

\begin{figure}[b!]
  \centering
  \subfloat[]{
  \begin{tikzpicture}
  \footnotesize

  \begin{axis}[
    width=.5\textwidth,
    height=.22\textheight,
    xmin=.015, xmax=.305,
    ymin=17, ymax=35,
    axis x line*=top,
    hide y axis,
    xtick={.02,.04,.06,.08,.10,.12,.14,.16,.18,.20,.22,.24,.26,.28,.30},
    xticklabel style={rotate=90},
    xticklabels={0.0010,0.0023,0.0037,0.0054,0.0070,0.0089,0.0109,0.0129,0.0149,0.0172,0.0197,0.0222,0.0247,0.0272,0.0297},
    xlabel={$K/N$ ($K$-AHS parameter)},
    xlabel near ticks,
    ]
  \end{axis}
  
  \begin{axis}[
    width=.5\textwidth,
    height=.22\textheight,
    xmin=.015, xmax=.305,
    ymin=17, ymax=35,
    xtick={.02,.04,.06,.08,.10,.12,.14,.16,.18,.20,.22,.24,.26,.28,.30},
    xticklabel style={rotate=90},
    xticklabels={0.02,0.04,0.06,0.08,0.10,0.12,0.14,0.16,0.18,0.20,0.22,0.24,0.26,0.28,0.30},
    ytick={10,15,20,25,30,35,40,45,50,55,60},
    yticklabels={10,15,20,25,30,35,40,45,50,55,60},
    xlabel={$M/N$},
    xlabel near ticks,
    ylabel={PSNR},
    ylabel near ticks,
    xmajorgrids,
    ymajorgrids,
    legend entries={K-AHS (Haar), K-AHS (CDF97), CS (Haar), CS (CDF97)},
    legend style={font=\scriptsize,at={(0.45,1.4)},
      anchor=south,legend columns=-1}]

  \addplot [
  color=blue,
  solid
  ]
  plot [error bars/.cd, y dir = both, y explicit]
  coordinates{
    (0.0200,19.7926) +- (0,0.1551) (0.0400,21.5604) +- (0,0.0765) (0.0600,22.5408) +- (0,0.1144) (0.0800,23.5822) +- (0,0.0859) (0.1000,24.1956) +- (0,0.0735) (0.1200,25.0472) +- (0,0.0850) (0.1400,25.6031) +- (0,0.0759) (0.1600,25.9811) +- (0,0.0933) (0.1800,26.3330) +- (0,0.1382) (0.2000,27.1442) +- (0,0.0760) (0.2200,27.4753) +- (0,0.1448) (0.2400,27.8124) +- (0,0.1107) (0.2600,28.1856) +- (0,0.0556) (0.2800,28.5129) +- (0,0.0801) (0.3000,28.7634) +- (0,0.0931) 
  };
  
  \addplot [
  color=black,
  solid
  ]
  plot [error bars/.cd, y dir = both, y explicit]
  coordinates{
    (0.0200,20.8300) +- (0,0.3225) (0.0400,23.1701) +- (0,0.0925) (0.0600,24.3013) +- (0,0.0798) (0.0800,25.4533) +- (0,0.1674) (0.1000,26.3122) +- (0,0.0983) (0.1200,27.3011) +- (0,0.0869) (0.1400,27.8587) +- (0,0.1848) (0.1600,28.4099) +- (0,0.1144) (0.1800,28.9862) +- (0,0.0982) (0.2000,29.7772) +- (0,0.1537) (0.2200,30.2816) +- (0,0.1026) (0.2400,30.7003) +- (0,0.0869) (0.2600,31.1280) +- (0,0.1082) (0.2800,31.4429) +- (0,0.1526) (0.3000,31.7555) +- (0,0.1237) 
  };

  \addplot [
  color=blue,
  dashed
  ]
  plot [error bars/.cd, y dir = both, y explicit]
  coordinates{
    (0.0200,18.2753) (0.0400,20.0324) (0.0600,21.3345) (0.0800,22.2587) (0.1000,23.1630) (0.1200,23.9450) (0.1400,24.6633) (0.1600,25.2402) (0.1800,25.8285) (0.2000,26.4500) (0.2200,26.9871) (0.2400,27.5221) (0.2600,28.0401) (0.2800,28.5430) (0.3000,28.9495) 
  };
  
  \addplot [
  color=black,
  dashed
  ]
  plot [error bars/.cd, y dir = both, y explicit]
  coordinates{
    (0.0200,18.3973) (0.0400,20.6110) (0.0600,22.3242) (0.0800,23.3894) (0.1000,24.4609) (0.1200,25.4214) (0.1400,26.2217) (0.1600,26.9497) (0.1800,27.7378) (0.2000,28.3791) (0.2200,29.0449) (0.2400,29.7075) (0.2600,30.2256) (0.2800,30.7907) (0.3000,31.3146) 
  };

\end{axis}
\end{tikzpicture}
}  
  
  \subfloat[]{\includegraphics[width=.24\textwidth]{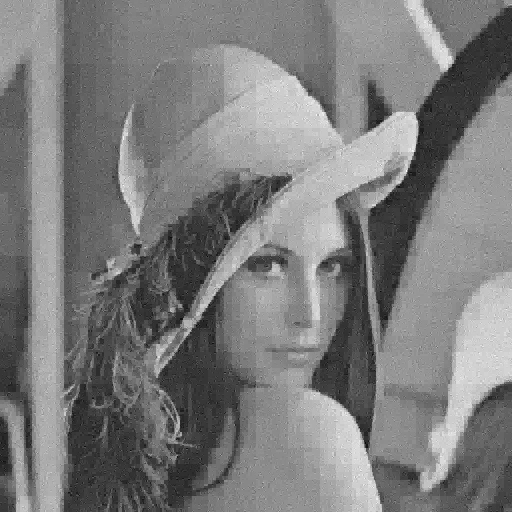}}
  \hfill
  \subfloat[]{\includegraphics[width=.24\textwidth]{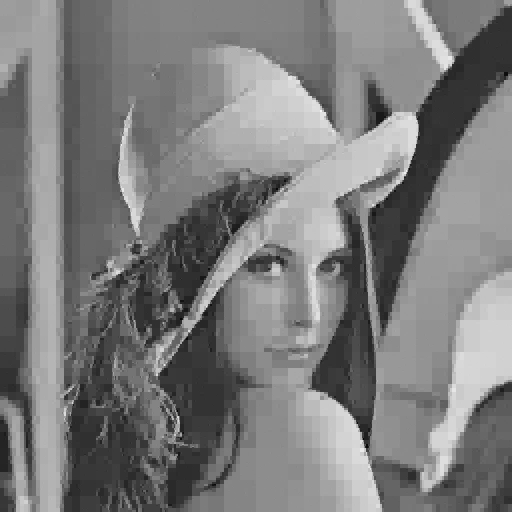}}

  \subfloat[]{\includegraphics[width=.24\textwidth]{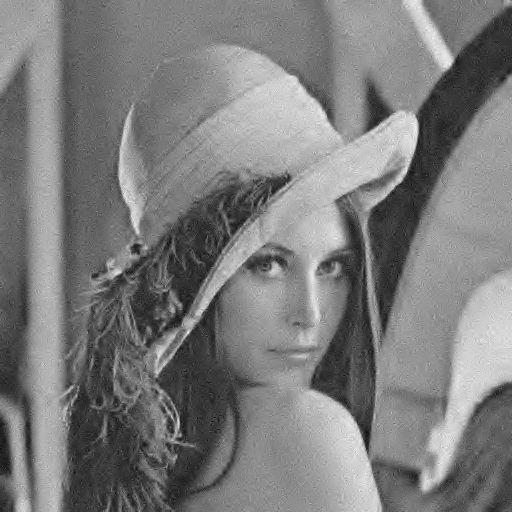}}
  \hfill
  \subfloat[]{\includegraphics[width=.24\textwidth]{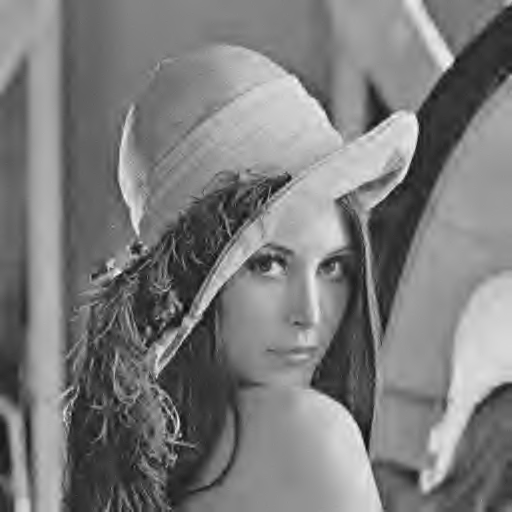}}
  
  \caption{Natural image sensing performance comparison between $K$-AHS and $\ell_1$-based CS for test image \textit{Lena}. (a) The PSNR dependent on the relative number of measurements. (b) CS reconstruction from $M=0.2N$ random noiselet measurements, Haar basis, PSNR: 26.45. (c) $K$-AHS reconstruction from $M=0.2N$ adaptive measurements, Haar basis, PSNR: 27.15. (d) CS reconstruction from $M=0.2N$ random noiselet measurements, CDF97 basis, PSNR: 28.38. (e) $K$-AHS reconstruction from $M=0.2N$ adaptive measurements, CDF97 basis, PSNR: 29.71. For visualization, reconstructed images were clipped to [0, 255] where necessary.}
  \label{fig:lena_results}
\end{figure}

\begin{figure}[b!]
  \centering
  \subfloat[]{
  \begin{tikzpicture}
  \footnotesize

  \begin{axis}[
    width=.5\textwidth,
    height=.22\textheight,
    xmin=.015, xmax=.305,
    ymin=17, ymax=30,
    axis x line*=top,
    hide y axis,
    xtick={.02,.04,.06,.08,.10,.12,.14,.16,.18,.20,.22,.24,.26,.28,.30},
    xticklabel style={rotate=90},
    xticklabels={0.0010,0.0023,0.0037,0.0054,0.0070,0.0089,0.0109,0.0129,0.0149,0.0172,0.0197,0.0222,0.0247,0.0272,0.0297},
    xlabel={$K/N$ ($K$-AHS parameter)},
    xlabel near ticks,
    ]
  \end{axis}
  
  \begin{axis}[
    width=.5\textwidth,
    height=.22\textheight,
    xmin=.015, xmax=.305,
    ymin=17, ymax=30,
    xtick={.02,.04,.06,.08,.10,.12,.14,.16,.18,.20,.22,.24,.26,.28,.30},
    xticklabel style={rotate=90},
    xticklabels={0.02,0.04,0.06,0.08,0.10,0.12,0.14,0.16,0.18,0.20,0.22,0.24,0.26,0.28,0.30},
    ytick={10,15,20,25,30,35,40,45,50,55,60},
    yticklabels={10,15,20,25,30,35,40,45,50,55,60},
    xlabel={$M/N$},
    xlabel near ticks,
    ylabel={PSNR},
    ylabel near ticks,
    xmajorgrids,
    ymajorgrids,
    legend entries={K-AHS (Haar), K-AHS (CDF97), CS (Haar), CS (CDF97)},
    legend style={font=\scriptsize,at={(0.45,1.4)},
      anchor=south,legend columns=-1}]

  \addplot [
  color=blue,
  solid
  ]
  plot [error bars/.cd, y dir = both, y explicit]
  coordinates{
    (0.0200,19.3576) +- (0,0.0984) (0.0400,20.7959) +- (0,0.0551) (0.0600,21.5018) +- (0,0.1086) (0.0800,22.3816) +- (0,0.1059) (0.1000,22.8807) +- (0,0.0893) (0.1200,23.6085) +- (0,0.0374) (0.1400,23.9167) +- (0,0.0929) (0.1600,24.2899) +- (0,0.0700) (0.1800,24.5867) +- (0,0.0596) (0.2000,25.2299) +- (0,0.0427) (0.2200,25.5204) +- (0,0.0656) (0.2400,25.8223) +- (0,0.0446) (0.2600,26.0842) +- (0,0.0598) (0.2800,26.3167) +- (0,0.0559) (0.3000,26.4914) +- (0,0.1811) 
  };

  \addplot [
  color=black,
  solid
  ]
  plot [error bars/.cd, y dir = both, y explicit]
  coordinates{
    (0.0200,19.9949) +- (0,0.3960) (0.0400,21.6658) +- (0,0.2435) (0.0600,22.4610) +- (0,0.1268) (0.0800,23.4941) +- (0,0.0859) (0.1000,23.9770) +- (0,0.1124) (0.1200,24.7733) +- (0,0.0670) (0.1400,25.1509) +- (0,0.1130) (0.1600,25.5334) +- (0,0.0644) (0.1800,25.9075) +- (0,0.0607) (0.2000,26.5805) +- (0,0.0410) (0.2200,26.8768) +- (0,0.0523) (0.2400,27.1926) +- (0,0.0405) (0.2600,27.4676) +- (0,0.1024) (0.2800,27.7432) +- (0,0.0691) (0.3000,27.9835) +- (0,0.0425) 
  };

  \addplot [
  color=blue,
  dashed
  ]
  plot [error bars/.cd, y dir = both, y explicit]
  coordinates{
    (0.0200,17.9749) (0.0400,19.2248) (0.0600,20.2702) (0.0800,21.0668) (0.1000,21.7512) (0.1200,22.3772) (0.1400,22.8703) (0.1600,23.3876) (0.1800,23.8848) (0.2000,24.3138) (0.2200,24.7712) (0.2400,25.1863) (0.2600,25.6655) (0.2800,26.0282) (0.3000,26.4848) 
  };
  
  \addplot [
  color=black,
  dashed
  ]
  plot [error bars/.cd, y dir = both, y explicit]
  coordinates{
    (0.0200,18.4693) (0.0400,19.5588) (0.0600,20.6135) (0.0800,21.4448) (0.1000,22.2839) (0.1200,22.8573) (0.1400,23.5437) (0.1600,24.0911) (0.1800,24.6246) (0.2000,25.1024) (0.2200,25.6114) (0.2400,26.1049) (0.2600,26.6155) (0.2800,26.9867) (0.3000,27.4789)
  };

\end{axis}
\end{tikzpicture}
}

  \subfloat[]{\includegraphics[width=.24\textwidth]{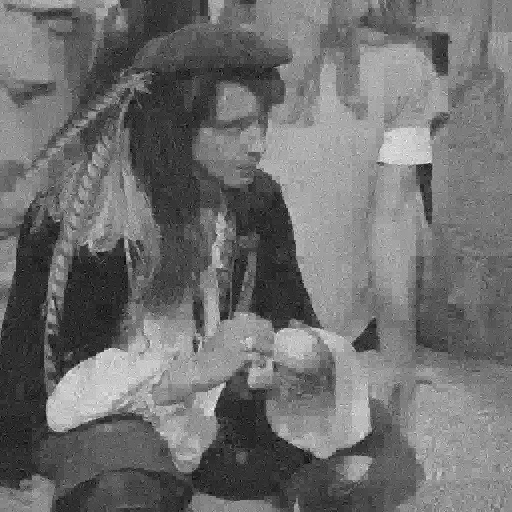}}
  \hfill
  \subfloat[]{\includegraphics[width=.24\textwidth]{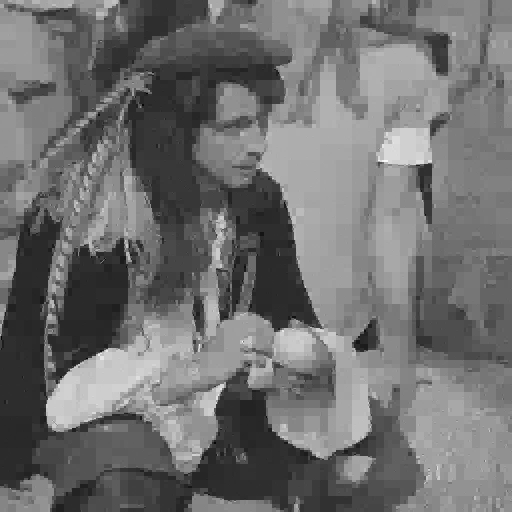}}

  \subfloat[]{\includegraphics[width=.24\textwidth]{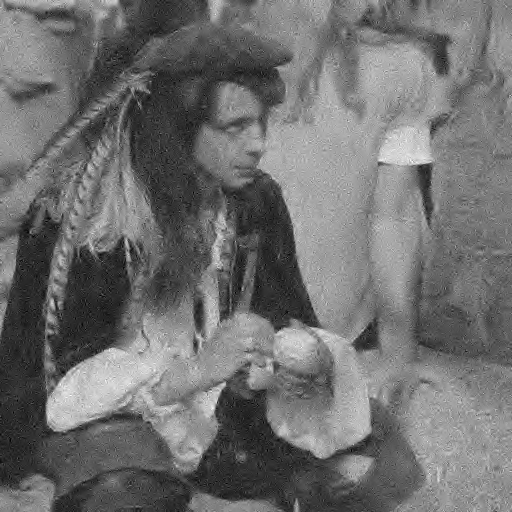}}
  \hfill
  \subfloat[]{\includegraphics[width=.24\textwidth]{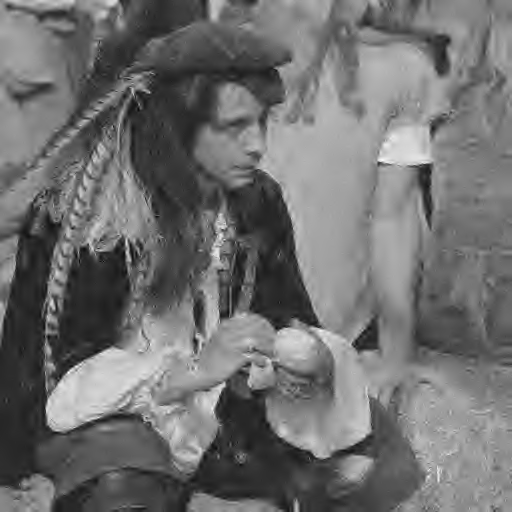}}
  \caption{Natural image sensing performance comparison between $K$-AHS and $\ell_1$-based CS for test image \textit{Pirate}. (a) The PSNR dependent on the relative number of measurements. (b) CS reconstruction from $M=0.2N$ random noiselet measurements, Haar basis, PSNR: 24.31. (c) $K$-AHS reconstruction from $M=0.2N$ adaptive measurements, Haar basis, PSNR: 25.26. (d) CS reconstruction from $M=0.2N$ random noiselet measurements, CDF97 basis, PSNR: 25.10. (e) $K$-AHS reconstruction from $M=0.2N$ adaptive measurements, CDF97 basis, PSNR: 26.57. For visualization, reconstructed images were clipped to [0, 255] where necessary.}
  \label{fig:pirate_results}
\end{figure}

\figurename s \ref{fig:cameraman_results}, \ref{fig:lena_results}, and \ref{fig:pirate_results} illustrate sensing results for the test images \textit{Cameraman}, \textit{Lena}, and \textit{Pirate}.

For each image, \figurename\ \ref{fig:cameraman_results}(a), \ref{fig:lena_results}(a), and \ref{fig:pirate_results}(a) illustrate the rate distortion analysis showing reconstruction accuracy as measured by the peak signal-to-noise ratio (PSNR) as a function of the number of collected measurements. Each curve corresponds to one of the four compressive imaging variants described above. Both approaches, $K$-AHS and CS, achieve consistently higher PSNR with the CDF97 wavelet basis than with the Haar wavelet basis. This can be explained by the fact that natural images have generally sparser representations by smooth CDF97 basis functions than by ternary, discontinuous Haar basis functions. For measurements up to $25$\% of the number of dimensions $N$ (usually $M \ll N$), the PSNR of $K$-AHS reconstructions is higher than the PSNR of CS reconstructions for both Haar and CDF97 wavelets. That difference is larger with the CDF97 basis than with the Haar basis. The reason might be that noiselets and Haar wavelets have minimal mutual coherence \cite{TuHu09} as opposed to the combination of noiselets and CDF97 wavelets for which the mutual coherence is small but not minimal \cite{PeLoSiCa14}. Therefore, it is more difficult for $K$-AHS to achieve higher reconstruction accuracy than CS. The larger the number of collected measurements, the smaller the PSNR difference between $K$-AHS and CS. For really large numbers of measurements, where $M \centernot{\ll} N$, CS reconstructions have higher PSNR than $K$-AHS reconstructions.

For each image, \figurename\ \ref{fig:cameraman_results}(b)-(c), \ref{fig:lena_results}(b)-(c), and \ref{fig:pirate_results}(b)-(c) illustrate CS and K-AHS reconstructions from $M = 0.2N$ measurements using the Haar wavelet domain. Each reconstructed image shows blocking artifacts, due to the discontinuity of the Haar wavelet basis. While both approaches restore edges and contours satisfactory, CS seems slightly more accurate at image regions containing small luminance variations. On the other hand, CS reconstructions suffer considerably from high frequency noise which is evenly distributed over the entire image and likely causing the inferior PSNR. $K$-AHS shows at some image regions slightly coarser block structures than CS but recovers overall homogeneous image regions more accurately. Furthermore, $K$-AHS does not suffer from high frequency noise.

For each image, \figurename\ \ref{fig:cameraman_results}(d)-(e), \ref{fig:lena_results}(d)-(e), and \ref{fig:pirate_results}(d)-(e) illustrate CS and K-AHS reconstructions from $M = 0.2N$ measurements using the CDF97 wavelet domain. In accordance with the rate distortion analysis, the images reconstructed in the CDF97 wavelet domain look, for both approaches, visually more pleasant than the images reconstructed in the Haar wavelet domain. Some contours of the $K$-AHS reconstructions show minor ringing artifacts whereas image regions with constant luminance and small luminance variation are more accurately recovered compared to CS. Again, images reconstructed by CS suffer from evenly distributed high frequency noise.

\begin{figure}[b!]
  \centering
  
  \input{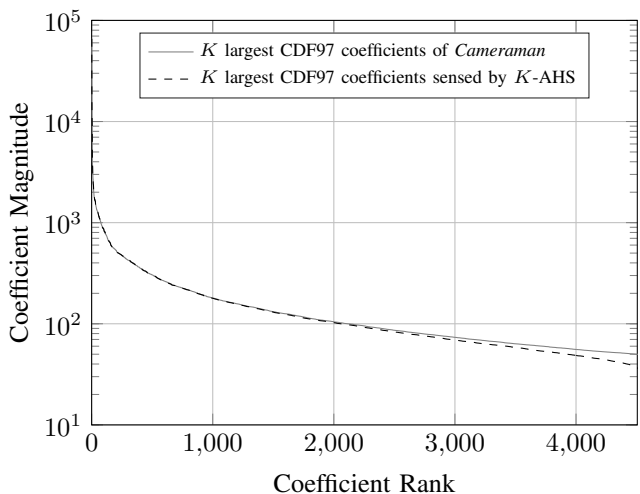}

  \caption{Comparison between the $K$ largest coefficients of image \textit{Cameraman} subject to the CDF97 wavelet basis, and the $K$ largest coefficients sensed by $K$-AHS. User parameter $K=4506$ ($M = 0.2N$).}
  \label{fig:captured_coefficients}
\end{figure}

For the image \textit{Cameraman}, \figurename\ \ref{fig:captured_coefficients} illustrates the magnitude of the $K$ largest coefficients in the CDF97 wavelet domain, as well as the $K$ largest coefficients that are sensed by $K$-AHS, where $K=4506$ ($M = 0.2N$). It can be seen that $K$-AHS collects a considerable number of the most significant CDF97 coefficients. The particular number of matches varies depending on the random permutation of the basis vectors. Over $1000$ runs with different random permutations, the average number of the most significant coefficients that are identified is $454.90$ (with a standard deviation of $188.02$). Although not all of the $K$ largest coefficients are identified, those coefficients found, have only a small deviation from the optimal ones.

\subsubsection{Spatial Sensing Maps}
\label{sec:spatial_sensing_maps}

\begin{figure}[b!]
  \subfloat[]{\includegraphics[width=.24\textwidth]{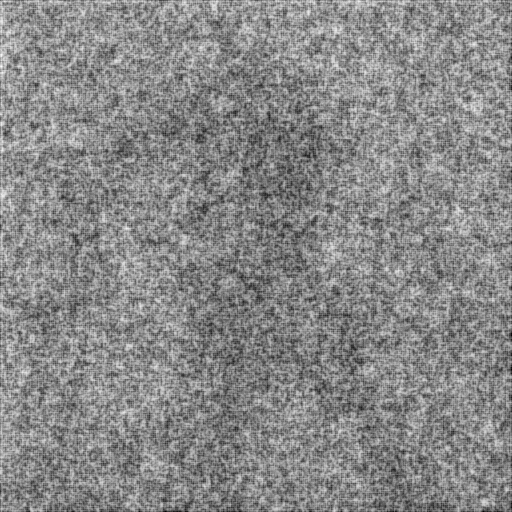}}
  \hfill
  \subfloat[]{\includegraphics[width=.24\textwidth]{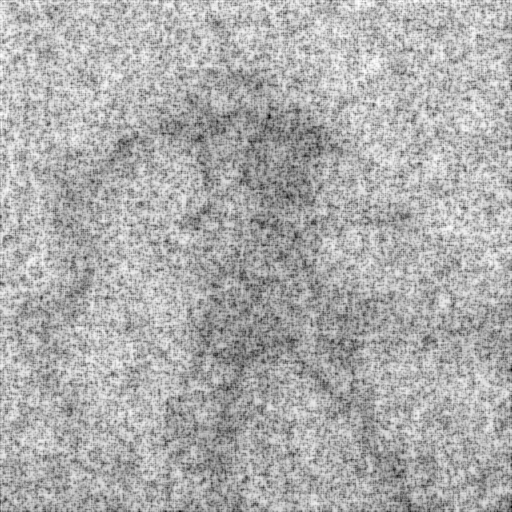}}

  \subfloat[]{\includegraphics[width=.24\textwidth]{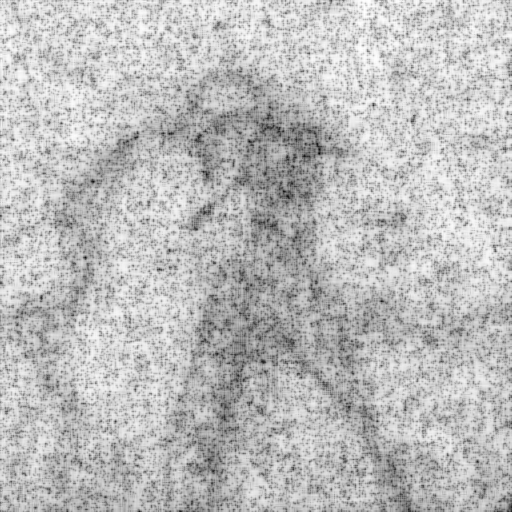}}
  \hfill
  \subfloat[]{\includegraphics[width=.24\textwidth]{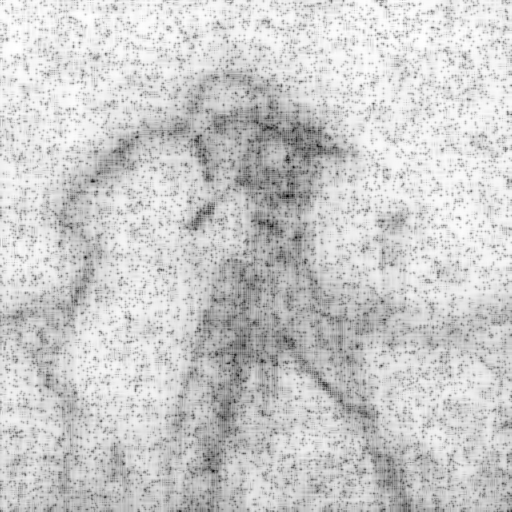}}
  
  \subfloat[]{\includegraphics[width=.24\textwidth]{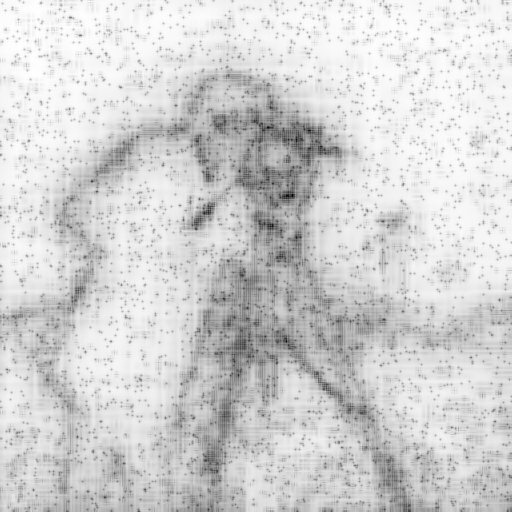}}
  \hfill
  \subfloat[]{\includegraphics[width=.24\textwidth]{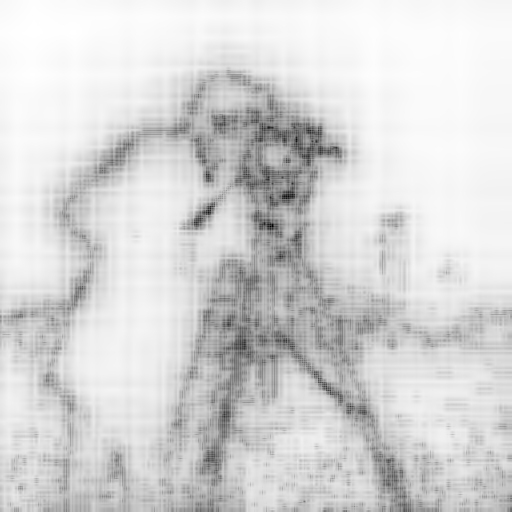}}
  
  \caption{Spatial sensing maps obtained while sampling the image \textit{Cameraman} by $K$-AHS in the CDF97 wavelet domain using $K=4095$. They indicate, for each level of the sensing tree, how intensively each region of the image is sensed by the $K$ ``winner sensing vectors'', i.e. where they are focusing the sensing load (see Section \ref{sec:spatial_sensing_maps}). (a) Initial level $L=5$. (b) Level $l=4$. (c) Level $l=3$. (d) Level $l=2$. (e) Level $l=1$. (f) Bottom level $l=0$. Each spatial map is normalized. White regions indicate minimal sensing activity, whereas black regions indicate maximal sensing activity.}
  \label{fig:spatial_sensing_maps}
\end{figure}

If the elements of the analysis basis are localized, as in the case of wavelets, the spatial regions at which $K$-AHS focuses its sensing load can be visualized. For each level processed by $K$-AHS, we identified the sensing vectors which provided the $K$ largest measurements, and replaced each of their entries by its absolute value. Subsequently, we calculated the sum of these $K$ rectified ``winner sensing vectors'' to obtain a spatial sensing map. This spatial sensing map indicates which image regions are sensed to which extent by the $K$ ``winner sensing vectors'' of the corresponding level. Since the $K$ winner determine in particular, by which branches of the sensing tree the sensing proceeds, they also determine, which regions shall be refined.
\figurename\ \ref{fig:spatial_sensing_maps} shows a sequence of spatial sensing maps from the initial level to the bottom level while sampling the image \textit{Cameraman} with $K=4095$. At the initial level, the spatial sensing map shows a rather broad and evenly distributed occurrence of regions. The image content is barely perceptible. As $K$-AHS descends to lower levels of the tree, the spatial sensing maps reveal more and more image structures. From the spatial map of the bottom level, the image content is well perceptible. Apparently, regions at which $K$-AHS focuses the sensing load are successively refined and lead to salient regions of the image such as distinct contours, edges and corners.

\section{Conclusion and Discussion}

In this paper, we have proposed $K$-AHS, a novel adaptive hierarchical procedure to sense sparse and compressible signals. As opposed to Compressed Sensing (CS), where non-adaptive measurements are collected by random sensing vectors, $K$-AHS adaptively selects sensing vectors from a collection depending on previous measurements of the signal. The sensing vectors are hierarchically organized in a sensing tree which is partially traversed by $K$-AHS during sampling. Each node of the tree represents a sensing vector, which is the sum of a subset of elements form the analysis basis $\mat{\Psi}$, which is chosen prior to sampling such that it provides a sparse representation of the signal.
When a node is visited, a linear measurement of the signal with this node-specific sensing vector is performed. Insignificant measurements cause the omission of subtrees and corresponding partitions of signal coefficients. Significant measurements, on the other hand, are iteratively refined by descending into their corresponding subtrees. Visited leaf nodes reveal signal coefficients in the sparse transform domain, whereas signal coefficients of unvisited leaf nodes are treated to be zero. Hence, the sparse representation of the signal is obtained without solving an optimization problem, a tremendous benefit over CS. Furthermore, with $K$-AHS (as opposed to CS) there are no pre-conditions demanding incoherence between the sensing vectors and the sparse synthesis transform.

We conducted a theoretical analysis which addresses the sensing quality of $K$-AHS in terms of detecting the $k$ most relevant signal coefficients. We provided a theorem, which states a general sufficient condition that guarantees to sense at least the optimal $k$-term approximation of the signal. Applying this condition, we investigated $K$-AHS sensing performance for three signal models as a function of their parameters. Experiments with synthetic signals of these models confirmed predictions according to our theoretical result. 

Based on experiments with natural images, we compared sensing performance of $K$-AHS to $\ell_1$-based CS in terms of image reconstruction accuracy as measured by PSNR using an orthogonal and a biorthogonal wavelet transform. Our general finding is, that for relevant numbers of measurements ($M \ll N$) $K$-AHS achieves better PSNR values than $\ell_1$-based CS.

The sensing vectors of the sensing tree $\{\vec{\varphi}_{l,j}\}$ can be pre-computed, which makes them instantly available. Consequently, adaptive sensing by $K$-AHS is essentially a process that selects and loads the requested sensing vectors on demand. This kind of pre-caching is the proposed default mode for $K$-AHS. Unlike other adaptive sensing approaches, it has the advantage that a requested sensing vector does not have to be computed specifically during sensing, which saves computational resources and time. On the other hand, pre-caching consumes memory of the order $\mathcal{O}(N^2)$. If, however, the amount of memory is limited, then each requested $\vec{\varphi}_{l,n}$ can be computed on demand taking the computational time of an analysis transform of an auxiliary vector with entries $1$ at the indices $(n-1)2^l+1, ..., n2^l$ and $0$ everywhere else (see Eq. (\ref{eq:sensing_vec_sum_leaves})). 
This can be fast nonetheless if $\mat{\Psi}$ is a fast transform such as the Discrete Cosine Transform (DCT) or the Fast Wavelet Transform (FWT). 
In a limited-memory-setting, CS sensing vectors are computed analogously on demand taking the computational time of a ``measurement transform'' (e.g. Fast Fourier Transform (FFT) or Fast Noiselet Transform (FNT)) of a randomly selected standard basis vector. Whichever mode is implemented, the additional computation time that is required by $K$-AHS at the sensing stage is due to sorting the measurements at the levels $L, L-1,\dots,1$. Note that this extra time is small: $\mathcal{O}(K \log K)$ for each level.

So far, we have considered the complete setting for $K$-AHS, where $\mat{\Psi}$ and $\overline{\mat{\Psi}}$ are bases. However, $K$-AHS can be applied analogously to the undercomplete setting, where $\mat{\Psi} \in \mathbb{R}^{Q \times N}\, (Q < N)$ and $\overline{\mat{\Psi}}^T$ is set to $\mat{\Psi}^\dagger$, the Moore-Penrose pseudoinverse of $\mat{\Psi}$.

As proposed in Section \ref{sec:extension_arbitrary_N_and_weighted_sums} sensing vectors of internal nodes could be generated by weighted sums rather than direct sums. In a future work, weights could be optimized and integrated into the tree composing process such that positively correlated non-zero coefficients have a higher probability to share common subtrees. To this end, typical samples from a signal class of interest could be analyzed numerically for statistical properties. Tuning the structure of the sensing tree and corresponding weights by a training stage might reduce the risk of falsely discarding significant coefficients.

\section*{Acknowledgement}

The research was funded by the DFG Priority Programme SPP 1527, grant number MA 2401/2-1.

\bibliographystyle{IEEEtran}
\bibliography{IEEEabrv,references}

\newpage
\begin{IEEEbiography}[{\includegraphics[width=1in,height=1.25in,clip,keepaspectratio]{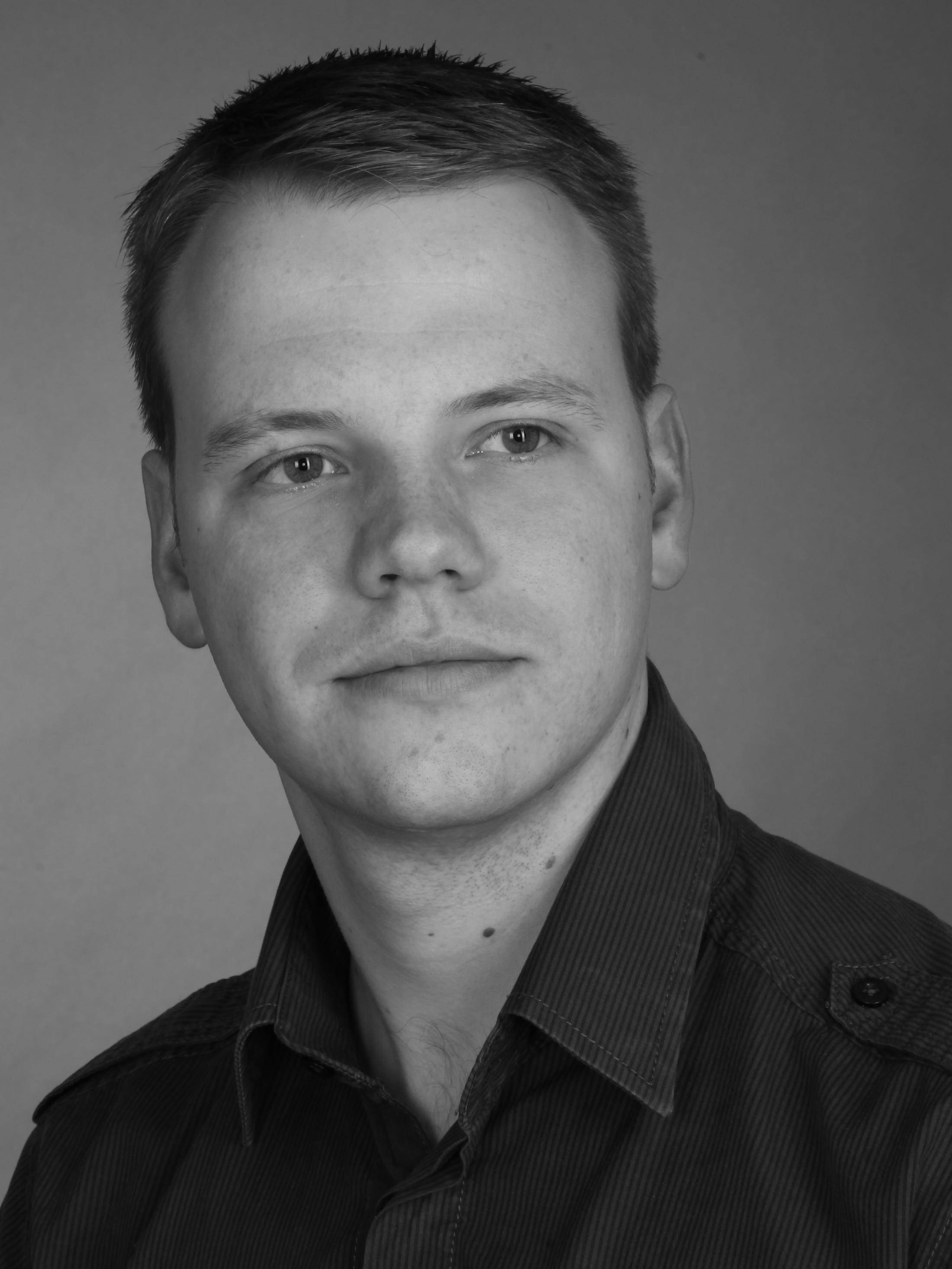}}]{Henry Sch\"utze}

studied Computer Science at the Brandenburg University of Technology, Cottbus, Germany and at the University of L\"ubeck, Germany. He graduated in 2011 with an MSc and is now research assistant at the Institute for Neuro- and Bioinformatics, University of L\"ubeck, where he pursues a PhD degree. His major research interests include Sparse Coding and Compressed Sensing.

\end{IEEEbiography}

\begin{IEEEbiography}[{\includegraphics[width=1in,height=1.25in,clip,keepaspectratio]{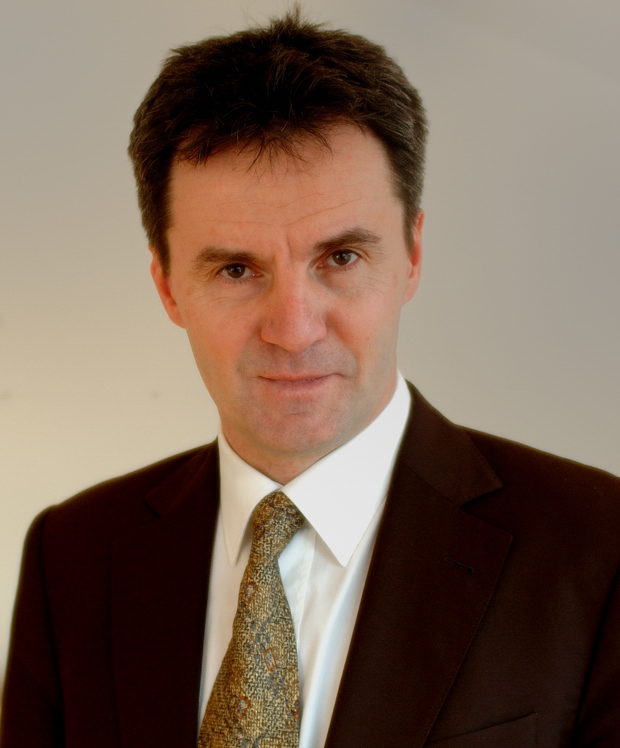}}]{Erhardt Barth}

received the Ph.D. degree in electrical and communications engineering from the Technical University of Munich, Germany. He is a Professor at the Institute for Neuro- and Bioinformatics, University of L\"ubeck, Germany, where he leads the research on human and machine vision. He has conducted research at the Universities of Melbourne and Munich, the Institute for Advanced Study in Berlin, and the NASA Vision Science and Technology Group in California.

\end{IEEEbiography}

\begin{IEEEbiography}[{\includegraphics[width=1in,height=1.25in,clip,keepaspectratio]{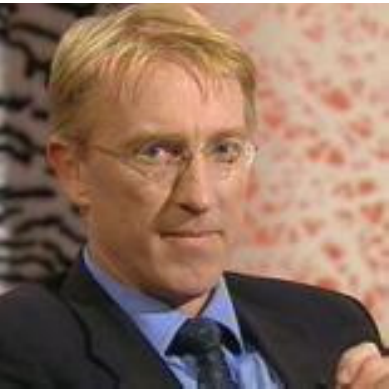}}]{Thomas Martinetz}

is full professor of computer science and director of the Institute for Neuro- and Bioinformatics at the University of L\"ubeck, Germany. He studied Physics at the Technical University of Munich, Germany and for his doctoral degree in Theoretical Biophysics joined the Beckman Institute for Advanced Science and Technology of the University of Illinois at Urbana-Champaign, USA. From 1991 to 1996 he led the project Neural Networks for automation control at the Corporate Research Laboratories of the Siemens AG in Munich. From 1996 to 1999 he was Professor for Neural Computation at the Ruhr-University of Bochum and head of the Center for Neuroinformatics.

\end{IEEEbiography}
\vfill

\end{document}